\documentstyle[epsf]{l-aa}
\def\Ti{$^{44}$Ti}

\def\thalf{$t_{1/2}$}
\def\tsh{t_{\rm sh}}
\begin{document}
\thesaurus{06
           (08.19.4;
            08.19.5 Cas~A;
            09.19.2; 
            02.14.1)}
 
\title{$^{\bf 44}$Ti: Its effective decay rate in young supernova
remnants, and its abundance in Cas~A}
 
\author{Y. Mochizuki \inst{1} \and
K. Takahashi \inst{2} \and
H.-Th. Janka \inst{2} \and
W. Hillebrandt \inst{2} \and 
R. Diehl \inst{3}
}
 
\offprints{Y.~Mochizuki (motizuki@postman. riken.go.jp)}
 
\institute{The Institute of Physical and Chemical Research (RIKEN), 
Hirosawa 2-1, Wako, Saitama 351-0198, Japan\ 
\and
Max-Planck-Institut f\"ur Astrophysik,
Karl-Schwarzschild-Stra\ss e 1, D-85740 Garching, Germany\ 
\and
Max-Planck-Institut f\"ur Extraterrestrische Physik,
Giessenbachstra\ss e 1, D-85740 Garching, Germany\
}
 
 
\maketitle
 
 
\begin{abstract}
 
Radioactive isotopes such as \Ti\ offer probably the most direct
probe into nucleosynthesis environments in the interior of 
exploding stars, when
the associated $\gamma$-ray activities in the explosion
remnant are detected and 
translated back to the
isotopic abundances  at the time of the explosion.
In this paper, we assert that the procedure may not necessarily 
be straightforward, at least in the case of \Ti, an orbital-electron
capture decay isotope.
Using the analytic model of McKee \& Truelove (1995)
for young supernova remnants, and
assuming the existence of overdense $^{56}$Fe-dominated clumps 
that contain also \Ti, we show
that a high degree of ionization may be 
caused by the reverse 
shock so that the electron-capture
rate of \Ti\ could be significantly reduced from its laboratory value.
When applied to Cas~A, this increases under certain
 conditions the 
current \Ti-activity by a factor $1.5 \sim 2.5$, which
yields a better compatibility between 
the COMPTEL
observation of the 1.16 MeV line activity associated with the \Ti\ 
decay and the SN model predictions of
the initial \Ti\ abundance.
This possibility is, however, subject to various uncertainties, and
in particular to the unknown properties and 
radial distribution of the clumps in the 
ejecta.
 
\keywords{
supernovae: general -- supernovae: individual: Cas~A --
ISM: supernovae remnants -- Nuclear reactions, nucleosynthesis, abundances
}
\end{abstract}
%
 
\section{Introduction}

Exploding massive stars, and supernovae in particular,
are known to be  major sites for the production of a large variety 
of elements heavier than carbon. 
One of the few available ways to study the physics 
in the deep interior of such stars is the
determination of the abundances of stable nuclides freshly produced
and ejected by the explosion. 
Infrared, optical, and X-ray spectroscopic measurements are capable of
determining
elemental abundances in the photosphere during different phases of 
the outburst. 
But the results from such observations are, in general, sensitive 
to the models
of line excitations in the photosphere, resulting in large 
correction factors to be applied before deducing
the isotopic yields at the time of the explosion.
In addition, results are hampered by the uncertainty of the 
optical depth, and by the possibility of 
heavy element condensation into 
dust shortly after the explosion, as was witnessed in 
SN 1987A. 

A more direct probe of massive-star interior physics is, in
principle, to investigate unstable nuclides and to
measure the $\gamma$-rays associated with their $\beta$ decays
after they have been ejected by the supernova explosion. 
For an optimum probe, the mean lifetime of such a radioactive isotope
should range from around a few weeks up to about
10$^6$ years. The lower limit is set by the requirement that the ejecta
should become optically thin to $\gamma$-rays in a few decay times,
and the upper limit by instrumental sensitivities of $\gamma$-ray 
telescopes
(the $\gamma$-ray flux from 
a trace isotope 
must exceed the instrumental noise level,
presently in the range of 10$^{-5}$ photons cm$^{-2}$s$^{-1}$, which
corresponds, for instance, to (several times)
$10^{-2}$ M$_\odot$ 
of an intermediate-mass isotope with 
a lifetime of $10^6$ y at the Galactic center or to 
$10^{-4}$ M$_\odot$ of the isotope 
in a supernova at a distance of
a few hundreds of pc).

Furthermore, whereas
short-lived isotopes will clearly trace individual events,
long-lived ones with mean lifetimes of the order of 10$^6$ y or longer
will reflect a superposition of different supernovae at different times, 
mixed with interstellar matter. Consequently clues to abundances in an
individual object are only very indirect.   
 
Only a few isotopes fulfill those constraints (see e.g. Diehl and 
Timmes 1998).
Most promising cases are found among the Fe
group elements, primarily because of their expected large abundances.  
The 0.847 and 1.238 MeV 
$\gamma$-ray lines from the $^{56}$Co $\rightarrow ^{56}$Fe decay
(half-life: \thalf\  = 77 d)
were detected from SN 1987A (Matz et al. 1988; Sandie et al. 1988;
 Mahoney et al. 1988; Rester et al. 1988; Teegarden et al. 1989).
There is also evidence for these decay lines from the unusually 
fast and bright
Type Ia supernova 1991T (Morris et al. 1995, 1997). 
The $^{57}$Co $\rightarrow ^{57}$Fe decay (\thalf\ = 272 d)
is another probe:
the 122 and 136 keV lines were detected
from SN 1987A (Kurfess et al. 1992; Clayton et al. 1992). 
Cases at the upper end of the favored radioactive lifetime range 
are $^{26}$Al (\thalf\ = 7.4 $\times 10^5$ y) and $^{60}$Fe
(\thalf\ = 1.5 $\times 10^6$ y).
The 1.809 MeV line from the $^{26}$Al decay has been
detected and mapped along the entire plane of the Galaxy (see 
review by Prantzos \& Diehl 1996).
If supernovae, rather than Wolf Rayet stars, were responsible 
for this $^{26}$Al,
the lines from the $^{60}$Fe $\beta^-$ decay would be
expected simultaneously with the $^{26}$Al decay, identifying 
a supernova origin 
(review by Diehl and Timmes 1998).
Instrumental sensitivity appears just at the borderline for this test.
 
In this paper, we focus our discussion on $^{44}$Ti, which decays 
with \thalf\ = 60 y 
(Ahmad et al. 1998; G\"orres et al. 1998; Norman et al. 1998;
Wietfeldt et al. 1999), making it ideal
for a study of inner-supernova physics within young supernova remnants.
$^{44}$Ti decays almost uniquely to the 2$^{nd}$ excited state 
of $^{44}$Sc, followed
immediately by the almost unique $\beta^+$ decay of $^{44}$Sc 
(\thalf\ = 4 h) to the 1.156 MeV excited state of $^{44}$Ca.
The 1.156 MeV de-excitation line has indeed been observed by the
COMPTEL telescope on the Compton Observatory
from Cas~A, a young supernova remnant with an estimated age of 
320 y  (Iyudin et al. 1994, 1997; Dupraz et al. 1997). 
The measured $\gamma$-ray flux is 
$\simeq(4 \pm 1)\times 10^{-5}$ photons /cm$^2$/s (Iyudin et al. 1997)
concordant with
an upper limit obtained by the OSSE instrument (The et al. 1996). 
With an adopted distance to Cas~A  of 3.4 kpc (Reed et al. 1995)
and the laboratory decay rate, the inferred initial mass of
$^{44}$Ti
is $\simeq$ $2 \times 10^{-4}$ M$_{\odot}$ (Iyudin et al. 1997; 
Woosley \& Diehl 1998). 
 
The current model predictions of the \Ti\ initial mass lie in an
 approximate
range of  $(6 \times 10^{-6} \sim 8 \times 10^{-5})$ M$_\odot$ 
for Type-II SNe (Woosley \& Weaver 1995; Thielemann et al. 1996), 
and of $(3 \sim 8) \times 10^{-5}$ M$_\odot$ for
Type-Ib SNe (Woosley et al. 1995), more or less strongly depending
on the progenitor masses.
Higher values up to $2 \times 10^{-4}$ M$_\odot$ were obtained
in some of the Type-II SN models, when 
progenitor masses
above 30 M$_\odot$ combine with high explosion energies
 (Woosley \& Weaver 1995),
and for a 20 M$_\odot$ (SN1987A) model star (Thielemann et al.~1996).
Barring the possibility that the progenitor of Cas~A happened 
to be such a star, 
one may conclude that COMPTEL observed significantly more
\Ti\ than expected (see, e.g., Fesen \& Becker 1991;
Hurford \& Fesen 1996 for discussions on the progenitor characteristics).
 
As far as its $\beta$-decay properties are concerned, 
\Ti\ is a very interesting trace isotope, because its decay mode
is pure orbital electron capture, which means that
fully ionized \Ti\ is stable.
Even  partial ionization of the innermost electrons
should lead to a considerably longer effective half-life (Mochizuki 1999).
Therefore, the question arises whether in supernova remnants \Ti\
could be highly ionized and thus more stable for a considerable period 
of time during the evolution.
In this case, it would be incorrect to use the
half-life measured in the laboratory, and initial abundances of \Ti\ 
as deduced from $\gamma$-ray intensities could be too high. 

However, there is no simple answer to this question and, as we shall
emphasize below, the thermodynamic history of a remnant
has to be known in detail before firm predictions can be made. 
On the other hand, it is interesting to speculate
whether the COMPTEL observations, which indicate an amount of
\Ti\ in Cas~A that appears higher than expected, reflect the effects
of an increased lifetime of the explosively-produced \Ti\ because of 
temporary and partial ionization. 

The primary aim of the present paper is to outline 
possible implications of (partial) ionization on the observable
$\gamma$-ray line flux from the decay of \Ti .
We shall present results obtained for a variety of different
conditions, based on a simple model for young supernova
remnants which, nonetheless, accounts for the features 
most relevant to this question. Of course, our main 
focus will be on Cas~A, the best studied case, and the
parameters of the model are chosen accordingly.
 
Section 2 contains an overview of observational aspects of 
SN explosions and remnants which are of relevance in the
context of this work.
In Sect.~3, we describe the employed model for young supernova remnants,
i.e., the analytic model of McKee \& Truelove (1995),
augmented by a description of the reverse shock 
interacting with denser
cloudlets (Sgro 1973; Miyata 1996). In addition, we describe 
the microphysics used in the model on which the 
calculation of the ``effective'' decay rate of \Ti\ is based.
In Sect.~4, we present our results for the time variation of
the \Ti\ decay rate, the \Ti\ abundance in young supernova remnants
and the associated $\gamma$-ray activities that can be measured.
As a specific example, we will preferentially use the case of Cas~A.
Summary and conclusions are given in Sect.~5.
 
\section{Young Supernova Remnants}
 
Studies of supernovae from the moment of the explosion up until
they dissolve and merge with the general interstellar medium has,
 in general,
been guided by observational opportunities as combined with  
model interpretations;
the interplay of different
physical processes, which vary spatially and rapidly 
within a young supernova remnant,
cannot be disentangled from measurements alone. Observational windows are
(from low to high energy radiation): Radio maps from 
electrons synchrotron-emitting
in magnetic-field structures around the shocked-gas region;
 infrared emission from
cool and hot dust within dense clumps embedded in the supernova
 remnant; optical
line emission from the edges of dense material embedded in the 
remnant's hot
plasma; X-ray emission from the hot, ionized, gas that has been 
shocked by the
forward blast wave and by the reverse shock traveling inward, 
respectively;
 and $\gamma$-rays from long-lived radioactivity such as \Ti .

Here we are mainly interested in the thermal history of the
radioactive \Ti\ after it has been ejected in a supernova
explosion. \Ti\ as well as other
iron-group elements are synthesized during the very early moments
of the explosion of a massive star 
in the layers adjacent to the nascent compact object (a neutron star or
black hole).
Therefore, observations of iron can be used to trace also \Ti .
However, not many good cases are known so far.  

SN~1987A is certainly the best studied case.
The large Doppler shifts of iron lines observed in the early spectra 
require that iron moves with velocities much higher than
some of the hydrogen which cannot 
be explained by spherically symmetric explosion models but indicates
that it is not homogeneously distributed in an expanding shell but
is found in clumpy structures. This conclusion is also
supported by the unexpectedly early detection of X- and 
$\gamma$-rays from the decay of radioactive elements 
(for summaries, see Woosley \& Weaver 1994, Nomoto et al.~1994),
which again suggests that these nuclei have been transported far out 
into the hydrogen-rich shells.
Other arguments in favor of this interpretation include
the smoothness of the light-curve  
(e.g., Woosley \& Weaver~1994, Nomoto et al.~1994 and references
therein), and the time-dependent features of the spectral lines
observed soon after the outburst (Utrobin et al. 1995), 
e.g., in the Bochum event (Hanuschik \& Dachs 1987; 
Phillips \& Heathcote 1989).  
In fact multi-dimensional supernova simulations 
(e.g., Herant et al. 1992, 1994; Shimizu et al. 1994; Shimizu 1995;
 Burrows et al. 1995;
Janka \& M\"uller 1995, 1996)
have demonstrated that the surroundings of the newly-formed neutron
star are stirred by hydrodynamic instabilities and that 
inhomogeneities and 
clumpiness of the products of explosive nucleosynthesis are likely the 
consequence. However, it cannot be excluded that SN 1987A is a special
case and not comparable with, e.g., Cas~A.
 
\Ti\ $\gamma$-ray emission is expected from core-collapse supernovae 
in general (e.g., Woosley \& Weaver 1995; Thielemann et al. 1996),
 yet there is
a significant deficit in such $\gamma$-ray line sources in the 
Galaxy for the
inferred Galactic core-collapse supernova rate (Dupraz et al. 1997).
The \Ti\ detection for Cas~A (Iyudin et al. 1994; 1997), 
despite the low Fe abundance, therefore has 
given rise to speculations about its exceptional nature, particularly
because its optical and X-ray characteristics support the 
idea of asymmetries
and a peculiar circumstellar environment (see below, and 
Hartmann et al.~1997 and references therein).
It was suggested, for example, that the progenitor of Cas~A might
have been a rapidly spinning star, in which
case more $^{44}$Ti could have been synthesized (Nagataki et
al. 1998). 

Recently,
a second Galactic \Ti\ source has been reported (Iyudin et al. 1998).
Its alignment with an also recently discovered X-ray remnant 
(Aschenbach 1998) 
suggests that it is a very young and most nearby supernova remnant,
 with an
age around 680y and a distance of 200pc only. This object still
 provides a puzzle
because of the absence of radio and optical emission expected for such a 
nearby supernova. Yet, if confirmed, it will provide a unique 
opportunity for the
study of supernova-produced \Ti; in which case 
it is of some interest to see if 
the modified decay rate of 
ionized \Ti\ addressed in our paper could also be important.
 
The \Ti\ found in young supernova remnants is probably 
formed during the $\alpha$-rich freezeout from
(near) nuclear statistical
equilibrium (Woosley et al. 1973; Woosley \& Weaver 1995;
Woosley et al. 1995; Thielemann et al. 1996; Timmes et al. 1996;
Timmes \& Woosley 1997; The et al. 1998).
Whereas \Ti\ is obviously fully ionized at the time of explosion,
\Ti\ ions will recombine with electrons during the adiabatic 
cooling phase,
which accompanies the expansion of the exploded star,
to become neutral after some 1000 s (e.g., Nomoto et al.~1994),
a negligible timescale when compared with the age of a supernova remnant.
Therefore, during most of this early phase \Ti\ will decay with the
 laboratory rate.
The subsequent evolution of the remnant may however provide conditions
for its re-ionization, which is the subject of our modeling effort.

The evolution of young supernova remnants has been modeled extensively,
 and the
spherically-symmetric explosion into homogeneous interstellar
surroundings appears well-understood (McKee and Truelove 1995). 
A free expansion
phase is followed by an adiabatic blast-wave phase, where interaction
with surrounding material produces outward and inward moving shock waves
 leading
to bright X-ray and radio emission, yet being unimportant for the
 energetics
of the remnant. Later phases of significant slowing-down and radiative
 losses
of the expanding remnant lie beyond the early phase where \Ti\ still
 decays.
Young remnants such as Cas~A are generally understood as being somewhere
intermediate between free expansion and the second phase, commonly called 
``Sedov-Taylor'' phase.

The evolution of such idealized young supernova
remnants during the ejecta-dominated stage ($t < t_{\rm ST}$)
and the Sedov-Taylor phase ($t > t_{\rm ST}$) was described
in an analytical model by McKee \& Truelove (1995). 
At the interface between ejecta and ambient medium a contact
discontinuity occurs, separating the exterior region of
shocked and swept-up circumstellar gas behind the outward-moving 
blastwave from the supernova remnant interior. Inward from 
the discontinuity,
a reverse shock travels
through the expanding, cold supernova ejecta, heating up the 
interior gas. 
Bright X-ray emission results from the hot plasma on both sides of the
contact discontinuity, with higher temperature on the outside heated by
the blastwave ($\geq$ keV), 
as compared with reverse-shock heated gas on the inside 
($\leq$ keV; e.g. Vink et al. 1996). Although such a two-temperature
model appears adequate to describe the X-ray emission of many
supernova remnants, considerable uncertainty remains in detail.
For example, the blastwave shock rapidly heats the ions entering the
blastwave region, 
the thermalization times for the X-ray emitting electrons
may exceed 100 years, so that non-equilibrium models are needed for a
proper description of X-ray and radio emission.
The detailed physical conditions within supernova remnants are far from
being understood, although the general evolution follows these
 fairly simple
descriptions quite closely, and is controlled by a few parameters,
the explosion energy, the mass of ejecta, and the surrounding 
medium density.
According to the McKee \& Truelove (1995) model, 
an explosion energy in the 
$(1 \sim 3) \times 10^{51}\,{\rm erg}$ range
does not seem unreasonable in the case of Cas~A
for the possible ejecta mass of 
$(2 \sim 5)$ M$_\odot$ (Tsunemi et al. 1986; Vink et al. 1996)
if the surrounding ambient gas density is of the order of
20 cm$^{-3}$ (Tsunemi et al. 1986)
and the current blastwave radius is $(2 \sim 3)$ pc 
for an assumed distance of 3.4 kpc
(Anderson \& Rudnick 1995; Holt et al. 1994; Jansen et al. 1988;
Reed et al. 1995).
 
Although
the general appearance of supernova remnant images in the radio
and in X-rays is that of a large-scale shell configuration as expected
from the above model, additional prominent clumpy structures appear in  
some cases
(e.g., Anderson \& Rudnick 1995; Koralesky et al. 1998). 
This suggests that the model outlined so far is an
oversimplification as far as details are concerned. 
The gross radio and X-ray emissivity may not be very sensitive to 
such discrepancy, tracing the electron component in the vicinity of the 
shock region, but the bulk ejected mass may be inadequately 
represented by the
inferred electron densities and temperatures
 (e.g., Koralesky et al. 1998). 
Thus, for the Cas~A remnant,
prominent structures have been studied in their 
respective forms of optical
knots and filaments, ``quasi-stationary flocculi", ``fast-moving knots", 
and ``fast-moving flocculi" (e.g., van den Bergh \& Kamper 1983;
 Reed et al. 1995;
Peimbert \& van den Bergh 1971; Chevalier \& Kirshner 1977, 1978, 1979;
 Reynoso et al. 1997;
Lagage et al. 1996).
There is overwhelming evidence for dense structures embedded in 
tenuous material
within the entire remnant, and even outside the blastwave shock
 radius. Obviously
the explosion itself produces fragments of material, seen now 
as fast-moving knots
with their abundance patterns supporting an ejecta origin. 
These clumps might carry heavier elements preferentially as
suggested by observations of fast-moving Fe clumps early in 
SN explosions,
e.g. in
SN 1987A (e.g., Nomoto et al.~1994, Wooden~1997 and references therein),
but a connection between the instabilities and clumpiness early
after the supernova explosion and the fragments and ``bullets'' seen 
in the remnants has not been established yet.

Given all these uncertainties in the evolution of supernova remnants
we do not attempt to model a specific object, such as Cas~A, in
detail here. We rather shall investigate by means of an
admittedly very simple model, varying its parameters within
reasonable limits, the potential effects
of ionization on the \Ti\ abundance estimates obtained from  
$\gamma$-ray observations.

\section{The Model}

The remnant model employed in this work uses the analytic description
by McKee \& Truelove (1995) for the hydrodynamic evolution.
The ejecta with mass $M_{\rm ej}$ are assumed to be
cold and to expand with explosion energy $E_{\rm ej}$ into a
homogeneous ambient medium which has a hydrogen number density
$n_{\rm H0}$ and is composed of 10 hydrogen atoms per helium atom,
corresponding to a helium mass fraction of 28.6\%.
Inside the homogeneous ejecta, we assume the existence of denser 
clumps that contain a small fraction of the ejecta mass but most
of the produced \Ti. The treatment of the interaction of the 
reverse shock with these clumps and of the corresponding effects on
the \Ti\ decay are described below.
Ionization of \Ti\ is most efficiently induced by electrons.
If and how long a high degree of ionization is sustained depends on 
the reverse-shock characteristics and
the postshock evolution of the matter containing \Ti.
 
%
\subsection{A model for young supernova remnants}

If sufficiently
high temperatures happen to be
reached behind the reverse shock, $^{44}$Ti  
may become fully ionized, in which case its decay is
prevented until cooling due to the expansion of the gas leads to 
(partial) recombination of the innermost electrons.

To follow the radioactivity of the supernova remnant requires
knowledge of the evolution of the shocked ejecta. The analytical
model of McKee \& Truelove (1995) yields scaled relations for the
radius, velocity and postshock temperature of the reverse shock
as functions of time $t$. Also, the model provides the density of
the unshocked (homogeneous) ejecta at time $t$, which allows one
to calculate the postshock density from the density jump at the 
reverse shock. In order to describe the density history of each mass 
shell after the reverse shock has passed through it, we assume
that the shocked matter moves (approximately) with the same velocity
as the contact discontinuity. Knowing the motion of the latter, 
therefore, we can estimate the dilution of the shocked gas
by the expansion of the supernova ejecta.

The time-dependent analytical solutions are given by the
three parameters, $M_{\rm ej}$, $E_{\rm ej}$ and $n_{\rm H0}$. A
combination of values of these parameters is considered to yield
a suitable description  of a certain supernova remnant, if the model
gives a present-day radius and velocity of the blastwave that is
compatible with observational data. 
 
We extended the analytic remnant model of McKee \& Truelove (1995)
by assuming that the iron-group elements together with the explosively 
nucleosynthesized \Ti\ are concentrated
in overdense clumps of gas. These clumps are assumed to move outward
through the dominant mass of homogeneous supernova ejecta with high 
velocity. Since the clumps should contain only a minor 
fraction of the total
mass of the ejecta, the dynamics of the young supernova remnant as 
described by the McKee \& Truelove (1995) model is assumed not to 
be altered 
by their
presence. The radial position $r$ of the iron clumps within the ejecta
can be measured by the relative mass coordinate $q$ which is defined as
%
\begin{equation}
q \equiv\, {4\pi r^3\rho_{\rm ej}\over 3\,M_{\rm ej}}\, ,
\end{equation}
\noindent
where $\rho_{\rm ej}$ is the uniform background density of the
ejecta which expand homogeneously.

\subsection{Thermodynamic conditions just behind the reverse shock}
 
We now consider the properties of a clump of metal-rich matter
within the homogeneous ejecta.
Denoting the density enhancement factor of the assumed clump
relative to the surrounding ejecta by 
$\alpha_{\rm clmp}$, 
we obtain the clump density after the reverse shock has passed 
at time $t_{\rm sh}$ as
%
\begin{equation}
   \rho (\tsh) = 4\, \alpha_{\rm clmp}\, \rho_{\rm ej} (\tsh) \, ,
\end{equation}
\noindent
for an adiabatic index $\gamma = 5/3$. 
 
Correspondingly, the post-reverse shock temperature of the clump
is related to that of the surrounding uniform ejecta,
$T_{\rm ej}(\tsh)$, as
%
\begin{equation}
   T (\tsh) = 
\frac{\beta(\alpha_{\rm clmp})}{\alpha_{\rm clmp}}\,
\frac{\mu(\tsh)}{\mu_{\rm ej}}\,
T_{\rm ej}(\tsh)\, ,
\end{equation}
\noindent
which generalizes by the factor $\mu(\tsh)/\mu_{\rm ej}$ 
the result of Sgro (1975) to the case considered here, where
the chemical composition of the clump is different from that of
the surrounding ejecta as described by the corresponding
mean molecular weights $\mu(\tsh)$  and $\mu_{\rm ej}$, 
respectively.
$T_{\rm ej}(\tsh)$ is given by 
%
\begin{equation}
T_{\rm ej}(\tsh) = \frac{3 \mu_{\rm ej}}{16 k}\, 
{\widetilde v}_{\rm r}^{\,2}(\tsh)\, ,
\end{equation}
where
${\widetilde v}_{\rm r}$ is the 
reverse-shock velocity in the rest frame of
the unshocked ejecta (McKee \& Truelove 1995).
In Eq.~(3),
$\beta$ is the ratio of the pressure values in the regions behind the
reflected shock  and in 
front of it. The reflected shock
 is formed when the reverse 
shock hits a dense clump in the ejecta
(Sgro 1975; Miyata 1996).  The ratio $\beta$ 
is related to $\alpha_{\rm clmp}$ as 
%
\begin{equation}
 \alpha_{\rm clmp} = \beta\, \Bigl( 1 + \frac{1-\beta}{\sqrt{4 \beta +1}}
 \Bigr)^{-2}\, ,
\end{equation}
\noindent
for $\gamma = 5/3$ (Sgro 1975).
In particular, $\beta = 1$ 
if $\alpha_{\rm clmp} = 1$ (no clumpiness), and
$\beta/\alpha_{\rm clmp}$ appearing in Eq.~(3) is less than unity for 
$\alpha_{\rm clmp} > 1.$
 
\subsection{Post reverse-shock evolution}
 
We now follow the fate of the homogeneous ejecta and of the
matter in clumps after they have experienced the impact of  the 
the reverse shock at time $\tsh$.
 As we have mentioned earlier, we
assume that the shock itself would continue to
proceed inward through the cold ejecta as prescribed 
by the McKee \& Truelove (1995) model.
The thermodynamic and chemical
evolution  of the  clumps 
at  times $t \geq \tsh$\ is treated as described below.
 
Since we trace the clump material, it is preferable to introduce
Lagrangian particle abundances, defined by
%
\begin{equation}
  n \equiv \frac{{\rm d} N }{{\rm d} q}\, ,
\end{equation}
\noindent
where d$N$ is the number of particles at the mass coordinate $q$
in the interval d$q$.
 
\subsubsection{Ionic abundances}
%
At a given time $t (\geq \tsh)$, 
the number abundance $n_i$ of a nuclear species $k$
with charge $Z$ 
and $i$ orbital electrons
(i.e. $Z - i$ times ionized, and $n_k = \sum_{i \geq 0} n_i)$
follows
%
\begin{eqnarray}
{\frac{{\rm d} n_i}{{\rm d} t}} = 
- [\lambda_{\beta,i} + 
\lambda_{{\rm ion},i} + \lambda_{{\rm rec},i}] n_i \nonumber \\
 +\lambda_{{\rm ion},i+1} n_{i+1} + \lambda_{{\rm rec},i-1} n_{i-1}\, ,
\end{eqnarray}
\noindent
where  $\lambda_{\beta,i}(t),
\lambda_{{\rm ion},i}(t),$ 
and $\lambda_{{\rm rec},i}(t)$ are 
the nuclear $\beta$-decay, ionization
and recombination rates, respectively, of the ion.
For \Ti, $\lambda_\beta
\neq 0$, and thus 
$n(t>\tsh) < n(\tsh) =  
n(0) {\rm exp}( - \lambda_{\rm lab} \tsh ),$
where $\lambda_{\rm lab}$ is the \Ti\ decay rate in the cold ejecta,  
i.e. its laboratory value.
 
Since we assume that material which contains \Ti\ consists
dominantly of one single stable nuclide ($^{56}$Fe in practice),
the number of ionization electrons, $n_{\rm e}$,
is given by $\sum_{i \geq 0} (Z - i) n_i$ for that species.
In practice, we further assume that both $^{56}$Fe and \Ti\ are 
ionized to some degree in the reverse shock itself, 
which we cannot resolve.
Typically, we start our network calculations with ionic 
systems of 10 bound electrons,
This choice is made 
since our major concern is to see the reduction of the orbital-electron
capture rates in the domains of high shock temperatures. 
 
The reaction rates entering in Eq.~(7) are discussed in
Appendix A.
 
\subsubsection{Ion and electron temperatures}
%
The network requires the knowledge of the thermodynamic evolution 
of the shocked matter, and in particular of the electron temperature,
$T_{\rm e}$, which  defines
the Boltzmann distribution, and of the matter density $\rho$
(or the ion number density, $n_{\rm ion}$),
which enters in the absolute electron number density $n_{\rm e}$.
 
The decrease by expansion of the temperature $T(t)$
from $T(\tsh)$ of Eq.~(3) would roughly
be proportional to $[\rho(t)/\rho(\tsh)]^{2/3}$
(for $\gamma = 3/5$ and $t > \tsh)$.
Two considerations are due, however.
One concerns the possibility that both $T_{\rm e}$ and the  
ion temperature, $T_{\rm ion}$, may not necessarily be the same as $T$.
Secondly, some non-adiabatic effects may appear as the result of
 ionization and
recombination.
 
It is sometimes asserted that the collisionless equilibration
between $T_{\rm ion}$ and 
$T_{\rm e}$  would 
occur quickly, in which case $T_{\rm ion} =
T_{\rm e} = T$ already at $t = \tsh$.
Save this possibility,
the equipartition of the shock energy to ions  and electrons may
not be achieved at once.
The equilibration process through the Coulomb interaction is described
by a non-linear equation (Spitzer 1962),
%
\begin{equation} 
 \frac{{\rm d} T_{\rm e}}{{\rm d} t} =  \frac{T_{\rm ion} - 
 T_{\rm e}}{\tau_{\rm eq}}\, ,
\end{equation} 
\noindent
in terms of the equilibration time $\tau_{\rm eq}(t)$, which is
a function primarily of
$T_{\rm ion}$, $T_{\rm e}$ and the number densities of ions and 
electrons.
Assuming that $T_{\rm e} \ll T_{\rm ion}$ at $t = \tsh$, and that
the total kinetic energy
%
\begin{equation}
  \frac{3}{2} ( n_{\rm ion} + n_{\rm e} ) T
  = \frac{3}{2} ( n_{\rm ion} T_{\rm ion}
+ n_{\rm e} T_{\rm e} ) 
\end{equation}
\noindent
is constant during a short time interval $\delta t$, we solve Eq.~(8)
simultaneously with the abundance network equations
to determine the degree of ionization 
of a dominant species, and thus $n_{\rm e}$.
 
In order to include the non-adiabatic effect caused by
ionization and recombination,
we assume that the energy required for unbinding additional electrons
(net
energy loss) is taken from electrons only. 
Namely, after performing the network calculation at each time step and
solving Eq.~(8), we make a replacement
%
\begin{equation}
    T_{\rm e} \rightarrow
 T_{\rm e} 
 {\rm exp}\, \Bigl( - \frac{\delta n_{\rm e}}{n_{\rm e}}
 - \frac{2}{3}
 \frac{ \delta Q}{n_{\rm e} T_{\rm e}}\Bigr)\, ,
\end{equation}
\noindent
where $\delta n_{\rm e}$ and $\delta Q$ are the
change of $n_{\rm e}$ during $\delta t$ and
the corresponding net energy consumption. 
The equilibrium temperature $T$ is then
calculated from Eq.~(9).
 
Finally, $T(t + \delta t),
T_{\rm ion}(t + \delta t),$ and
$T_{\rm e}(t + \delta t)$ are obtained from the
current values by multiplying $[\rho(t +\delta t) /
\rho(t)]^{2/3}.$
 
\section{Results}
 
We present here our results, which we obtained with
 the model described above. They depend on
the following parameters:
$E_{\rm ej}, M_{\rm ej}, n_{\rm H0}$, $\alpha_{\rm clmp}$ and two 
values for $q$,  given as $q_0$ and $q_1$, which define the
lower and upper boundaries of the mass shell where the \Ti-containing
clumps are assumed to exist.
As we mentioned earlier, the values of the first three 
can be constrained in the case of Cas~A
from the observational analyses of
the blast-wave radius $R_{\rm b}$ and velocity $v_{\rm b}$.
Below  we consider mainly the parameter space which is
consistent with these constraints.
While we treat $\alpha_{\rm clmp}$ as a free parameter, a value of
the order of 10 seems to be reasonable as a first guess,
based on the density contrast which develops in hydrodynamic
instabilities in the envelope of the progenitor star during
the supernova explosion (Fryxell et al. 1991).
 
\begin{figure}
\epsfxsize=0.99\hsize
\epsffile{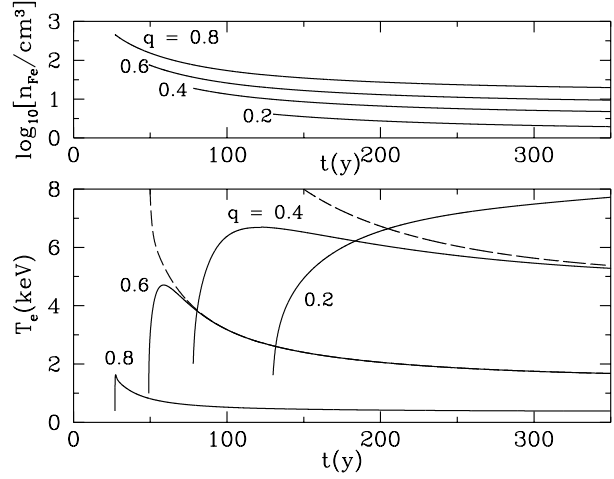}
\caption{Examples for the evolution with time $t$ (in years) of the
$^{56}$Fe number density $n_{\rm Fe}$
({\it top}), and of the electron temperature 
$T_{\rm e}$ ({\it bottom}) in the
$^{56}$Fe-dominated clumps at four different locations indicated  
by the  mass coordinate $q$.
The dashed lines represent the asymptotic behavior of the 
 equilibrium temperatures $T(t)$ (as described in Sect.~3.3.2)
at $q = 0.4$ and $0.6$.
The left edge of each solid curve corresponds to $t = \tsh$. For example,
the reverse shock took 50 y to reach $q = 0.6$.
The results are given for  the following set of parameter values:
$E_{\rm ej} = 3 \times 10^{51}$ erg,
$M_{\rm ej} = 3$ M$_\odot, n_{\rm H0} = 15$ cm$^{-3}$ and
$\alpha_{\rm clmp} = 10$
}
\end{figure}
%
\subsection{Thermodynamic evolution of post-shock clumps}
%
Figure 1 shows the evolution of the 
$^{56}$Fe
number density
$n_{\rm Fe}$  and of
the electron temperature $T_{\rm e}$ in the $^{56}$Fe-dominated
clumps at four different locations in the ejecta, given by
the corresponding mass coordinate $q$, after the reverse shock
has passed the clumps.
The post-shock expansion decreases the density as well as the
equilibrium  temperature,
while $T_{\rm e}$ first increases in its attempt to reach 
equilibrium between electrons and ions.
The results are plotted for the chosen set of parameter values 
quoted therewith.
 
\begin{figure}
\epsfxsize=0.99\hsize
\epsffile{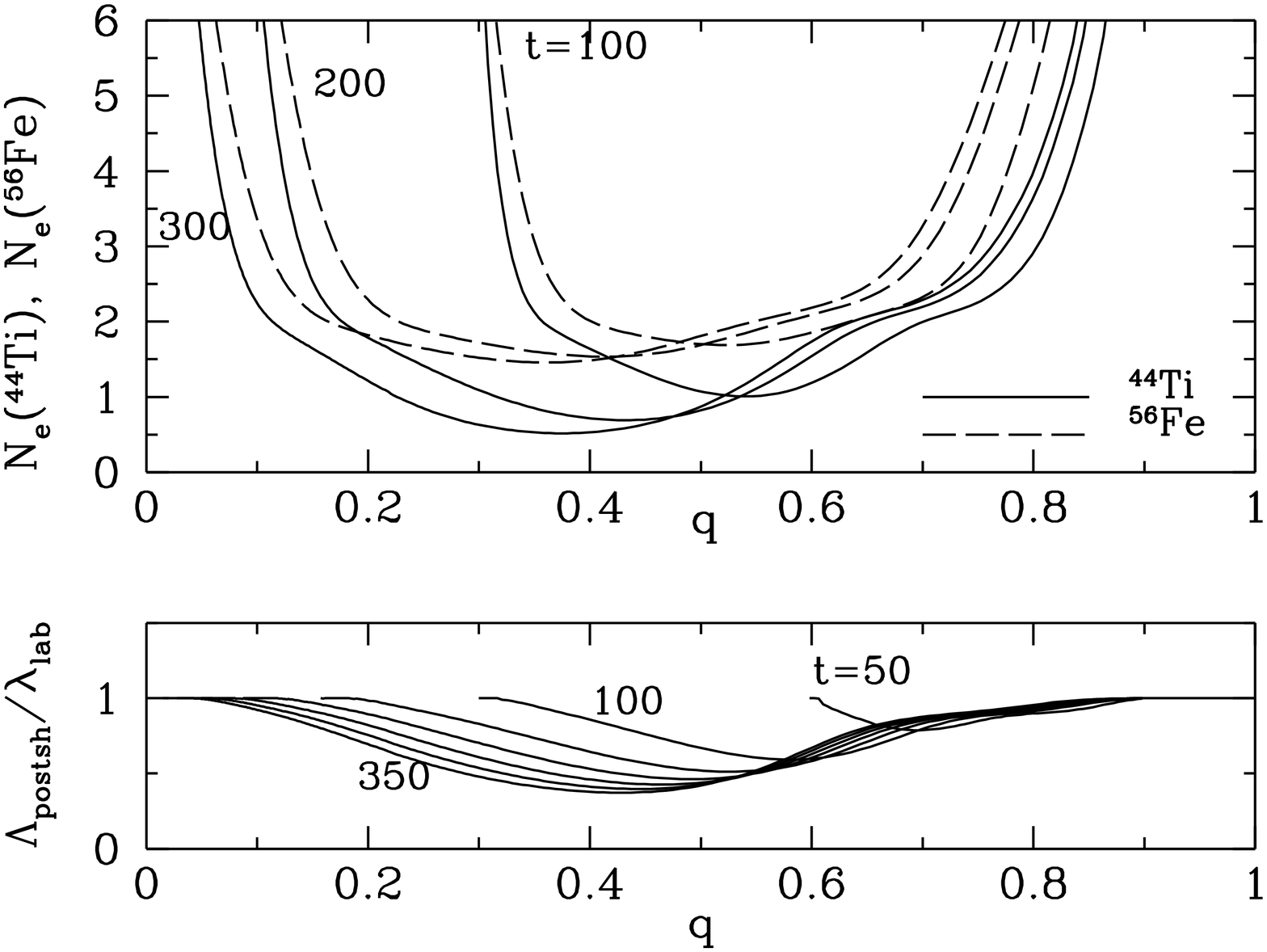}
\epsfxsize=0.99\hsize
\epsffile{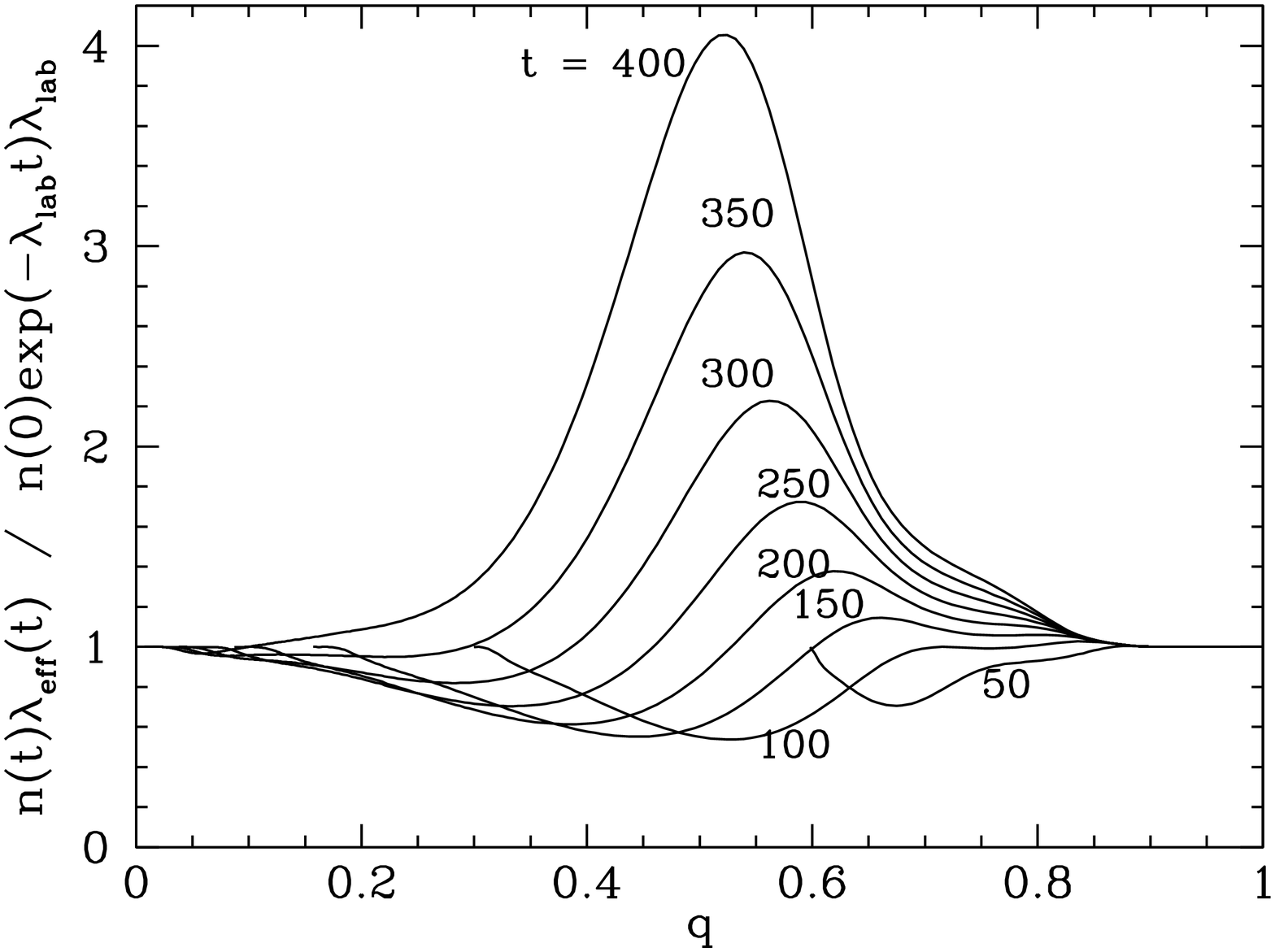}
\caption{{\it Top panel:} The average numbers  $N_{\rm e}$ of electrons
that are bound
to \Ti\ and to $^{56}$Fe at mass coordinate $q$ and
at time $t$ (in y) after the explosion.
{\it Middle:} Effective decay rates 
$\Lambda_{\rm postsh}$ of the shocked \Ti\ during the time span between
$\tsh$\
and $t$ [Eq.~(11)]
in units of the laboratory rate $\lambda_{\rm lab}$.
{\it Bottom:} The corresponding \Ti\ 
radioactivity observable at time $t$ (``age'') relative
to the case that assumes no reduction of the
$\beta$-decay rate.
The parameter values used here are the same as those for Fig.~1
}
\end{figure}
%
\begin{figure}
\epsfxsize=0.99\hsize
\epsffile{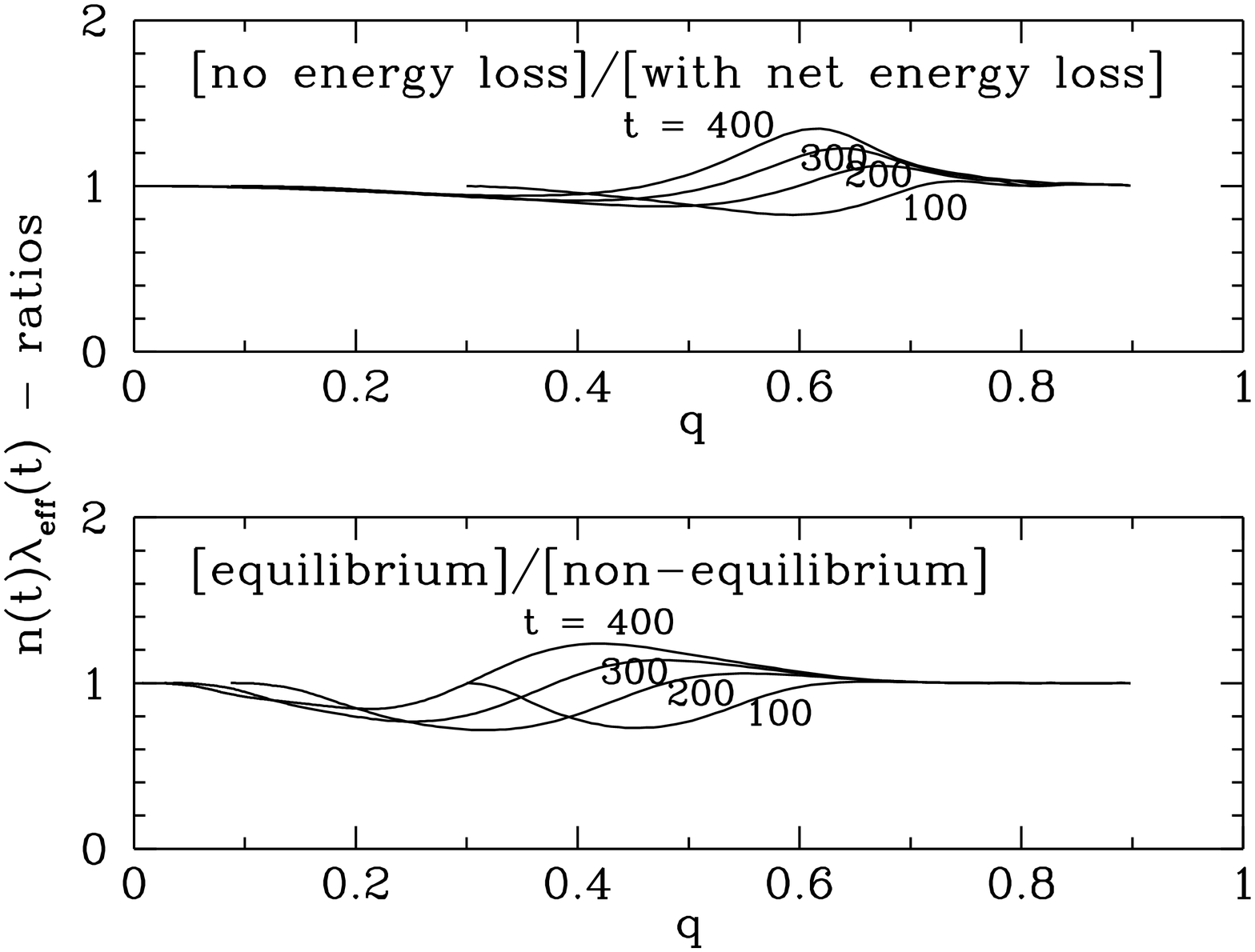}
\caption{\Ti\ 
 radioactivities 
at times $t$ (in y) in the clumps located at radial positions $q$,
relative to the radioactivity of the model with most detailed
input physics, when
the non-adiabaticity of the thermal evolution of the clumps
 is ignored ({\it upper panel}) and when
instantaneous equilibration between
$T_{\rm e}$ and $T_{\rm ion}$ 
(toward $T$) is assumed ({\it lower panel})
}
\end{figure}
%
\subsection{Effective decay rates of 
$^{44}${\rm Ti} and its radioactivity 
as an observable}
 
Let us now consider \Ti\ in a clump that is located anywhere in the
mass coordinate $0 \leq q \leq 1$ of the ejecta.
If \Ti\ is  near the surface (large $q$), it suffers from the reverse  
shock early. Since the electron
temperature is still relatively low 
(see Fig.~1),
there is  no possibility to reach a high degree of ionization (thus
an appreciable reduction of the decay rate).
On the other hand, \Ti\ embedded in clumps located
in the innermost region may become highly ionized
as the temperature is generally highest, resulting in the  smallest
\Ti\ decay rate.  Since
$\tsh$ is closer to $t$, the ``age'' of the remnant, however, the net
effect of the reduced decay rate appears weaker. Therefore,
the maximum effect of the retarded $\beta$-decay
is obtained  if \Ti\ is at intermediate values of $q$. 
This situation is depicted in the panels of Fig.~2.
 
The top panel of Fig.~2 shows the average numbers of electrons
 that are bound
to \Ti\ and to $^{56}$Fe at $t = 100, 200$ and 300 years 
after the explosion.
The increase of $N_{\rm e}$ toward the low side of $q$ indicates
 that the  time  
span between $\tsh$ and $t$ 
was not long enough for ionization.
At the high end of $q$-values, the effects of post-shock recombination  
are visible. 
When combined with $n_{\rm Fe}$ from Fig.~1, $N_{\rm e}(^{56}$Fe)
determines the total number of ionization electrons, $n_{\rm e}(t)$,
in a clump: 
$n_{\rm e} = [Z - N_{\rm e}(^{56}$Fe)] $n_{\rm Fe}$ with $Z = 26$.
 
The retardation of the \Ti\ decay by ionization is illustrated in 
the middle panel of Fig.~2. 
The effective decay rate in the post-shock period,
$\Lambda_{\rm postsh}$, is 
defined as the time-average
of $\lambda_{\rm eff}$ [given by Eq.~(18) in Appendix A]
between the time $\tsh$\ and a time $t > \tsh$.
It can be expressed as
%
\begin{equation}
     \Lambda_{\rm postsh} \equiv  - \frac{1}{t - \tsh}\,
      {\rm ln}\Bigl[ \frac{n(t)}{n(\tsh)} \Bigr]\, ,
\end{equation}
\noindent
where $n(t)$ is the \Ti\ abundance at mass coordinate $q$ and
time (age) $t$, and $n(\tsh)$
is the corresponding
 value at the shock impact time $\tsh$, which has been reduced
from its initial value $n(0)$ according to
$n(\tsh) =  n(0)
       {\rm exp}(- \lambda_{\rm lab} \tsh)$.
 
In order to estimate the observable consequences of the reduced
$\beta$-decay rates,
one has to recall that the measurable  $\gamma$-ray
activity per (normalized) unit mass of the remnant 
is the product of the current $\beta$-decay rate and the current
abundance, $n(t) \lambda_{\rm eff}(t)$.
 This introduces a nonlinear relation between 
the observable activity given 
in the bottom panel of Fig.~2 and the effective decay rates 
displayed above,
particularly in the absence of quick recombination.
 
In Fig.~3, we compare the results obtained with the input physics
as described in Sect.~3 to the cases where either the non-adiabaticity
of the equilibration process of electrons and ions owing to
the energy consumption by the ionization process is neglected
({\it upper panel}), or where instantaneous equilibration 
 (i.e., $T_{\rm e} = T_{\rm ion} = T$ always)
is assumed 
({\it lower panel}).
It can be seen that in both cases the trends of the 
``full'' model shown in Fig.~2, where both the non-adiabaticity
and the gradual equilibration process are taken into account,
are somewhat enhanced.
This can be understood by the fact that the omission
of either of the two effects leads to higher electron
temperatures and thus to a  higher degree  of ionization in the
clumps.
 
\subsection{$^{44}${\rm Ti} radioactivity in Cas~A}

The  \Ti\ 
radioactivity of the whole supernova remnant as 
observable at an age $t$, relative to the activity level that
would result in the absence of any retardation of 
the $\beta$-decay rate by ionization, is defined by the ratio
%
\begin{equation}
  {\cal A} \equiv 
\int_{q_0}^{q_1} n(q, t) \lambda_{\rm eff}(q, t) {\rm d}q
{\bigg /}  \Bigl[ N(0){\rm exp}( - \lambda_{\rm lab} t) \lambda_{\rm lab}
\Bigr]\, .
\end{equation}
\noindent
Here, $N(0)$ is the total number of \Ti\  nuclei in the $^{56}$Fe
clumps 
at $t = 0$.
For a remnant like Cas~A, which is in the transition phase from
the ejecta-dominated stage to the Sedov-Taylor phase,
the reverse shock has passed through most of the ejecta. 
Therefore, the factor ${\cal A}$ must be expected to be larger
than unity when the remnant is observed at this stage of its
evolution.
This implies that the initial abundance inferred from
 the current $\gamma$-ray activity due to the \Ti\ decay
is lower by a factor ${\cal A}^{-1}$ than the 
amount of produced \Ti\ that is estimated on the basis of the laboratory
decay rate $\lambda_{\rm lab}$.
 
\begin{table}
\caption{Illustration of the variability of the
\Ti\ radioactivity ratios ${\cal A}$ as defined by Eq.~(12)
with different values of the model parameters.
The tabulated results are at the time (age) of 320 y
after the explosion for an explosion energy of
$E_{\rm ej} = 1 \times 10^{51}$ erg and various combinations of
values for the ejecta mass $M_{\rm ej}$,
the hydrogen number density in the ambient medium, $n_{\rm H0}$,
the over-density factor of the clumps, $\alpha_{\rm clmp}$, and
the lower and the upper boundaries $q_0$ and $q_1$ 
of the mass shell 
 where the clumps are
assumed to exist.
$R_{\rm b}$ and $v_{\rm b}$ are the radius and velocity of the
blast-wave at that time as computed from the McKee \& Truelove (1995)
model for the given sets of $E_{\rm ej}, M_{\rm ej}$ and $n_{\rm H0}$
values
}
$$\vbox{\tabskip4pt \halign{#&#&#&#&#&#&#&# \cr
\noalign{\hrule}
\noalign{\smallskip}
\hfil $M_{\rm ej}$ & 
\, $n_{\rm H0}$ & \ $R_{\rm b}$ & \ $v_{\rm b}$ 
& \, $\alpha_{\rm clmp}$ & \ $q_0$ & \ $q_1$ & \ ${\cal A}$ \cr
 (M$_\odot$) & (cm$^{-3}$)
& (pc) & $(10^3$ km/s) & & & \cr
\noalign{\smallskip}
\noalign{\hrule}
\noalign{\smallskip}
\noalign{\smallskip}
\ 2 & \ 15 & 1.71 & \ 2.53 &
\ 5 & \hfil 0.4 & \hfil 0.6 & \hfil 1.47 \cr
\ 2 & \ 15 & 1.71 & \ 2.53 &
\ 10 & \hfil 0.0 & \hfil 0.5 & \hfil 1.12 \cr
\ 2 & \ 15 & 1.71 & \ 2.53 &
\ 10 & \hfil 0.4 & \hfil 0.6 & \hfil 1.55 \cr
\ 2 & \ 15 & 1.71 & \ 2.53 &
\ 10 & \hfil 0.2 & \hfil 0.8 & \hfil 1.27 \cr
\ 2 & \ 15 & 1.71 & \ 2.53 &
\ 20 & \hfil 0.4 & \hfil 0.6 & \hfil 1.34 \cr
\noalign{\smallskip}
\ 2 & \ 30 & 1.51 & \ 2.15 &
\ 5 & \hfil 0.4 & \hfil 0.6 & \hfil 1.92 \cr
\ 2 & \ 30 & 1.51 & \ 2.15 &
\ 10 & \hfil 0.0 & \hfil 0.5 & \hfil 1.30 \cr
\ 2 & \ 30 & 1.51 & \ 2.15 &
\ 10 & \hfil 0.4 & \hfil 0.6 & \hfil 1.77 \cr
\ 2 & \ 30 & 1.51 & \ 2.15 &
\ 10 & \hfil 0.2 & \hfil 0.8 & \hfil 1.43 \cr
\ 2 & \ 30 & 1.51 & \ 2.15 &
\ 20 & \hfil 0.4 & \hfil 0.6 & \hfil 1.35 \cr
\noalign{\smallskip}
\ 3 & \ 15 & 1.65 & \ 2.66 &
\ 5 & \hfil 0.4 & \hfil 0.6 & \hfil 1.27 \cr
\ 3 & \ 15 & 1.65 & \ 2.66 &
\ 10 & \hfil 0.0 & \hfil 0.5 & \hfil 1.04 \cr
\ 3 & \ 15 & 1.65 & \ 2.66 &
\ 10 & \hfil 0.4 & \hfil 0.6 & \hfil 1.27 \cr
\ 3 & \ 15 & 1.65 & \ 2.66 &
\ 10 & \hfil 0.2 & \hfil 0.8 & \hfil 1.13 \cr
%
%
\ 3 & \ 15 & 1.65 & \ 2.66 &
\ 20 & \hfil 0.4 & \hfil 0.6 & \hfil 1.18 \cr
\noalign{\smallskip}
\ 3 & \ 30 & 1.47 & \ 2.23 &
\ 5 & \hfil 0.4 & \hfil 0.6 & \hfil 1.55 \cr
\ 3 & \ 30 & 1.47 & \ 2.23 &
\ 10 & \hfil 0.0 & \hfil 0.5 & \hfil 1.13 \cr
\ 3 & \ 30 & 1.47 & \ 2.23 &
\ 10 & \hfil 0.4 & \hfil 0.6 & \hfil 1.38 \cr
\ 3 & \ 30 & 1.47 & \ 2.23 &
\ 10 & \hfil 0.2 & \hfil 0.8 & \hfil 1.22 \cr
%
%
\ 3 & \ 30 & 1.47 & \ 2.23 &
\ 20 & \hfil 0.4 & \hfil 0.6 & \hfil 1.20 \cr
\noalign{\smallskip}
\ 4 & \ 15 & 1.59 & \ 2.80 &
\ 5 & \hfil 0.4 & \hfil 0.6 & \hfil 1.14 \cr
\ 4 & \ 15 & 1.59 & \ 2.80 &
\ 10 & \hfil 0.0 & \hfil 0.5 & \hfil 1.00 \cr
\ 4 & \ 15 & 1.59 & \ 2.80 &
\ 10 & \hfil 0.4 & \hfil 0.6 & \hfil 1.15 \cr
\ 4 & \ 15 & 1.59 & \ 2.80 &
\ 10 & \hfil 0.2 & \hfil 0.8 & \hfil 1.06 \cr
%
%
\ 4 & \ 15 & 1.59 & \ 2.80 &
\ 20 & \hfil 0.4 & \hfil 0.6 & \hfil 1.11 \cr
\noalign{\smallskip}
\ 4 & \ 30 & 1.43 & \ 2.32 &
\ 5 & \hfil 0.4 & \hfil 0.6 & \hfil 1.33 \cr
\ 4 & \ 30 & 1.43 & \ 2.32 &
\ 10 & \hfil 0.0 & \hfil 0.5 & \hfil 1.05 \cr
\ 4 & \ 30 & 1.43 & \ 2.32 &
\ 10 & \hfil 0.4 & \hfil 0.6 & \hfil 1.22 \cr
\ 4 & \ 30 & 1.43 & \ 2.32 &
\ 10 & \hfil 0.2 & \hfil 0.8 & \hfil 1.12 \cr
%
%
\ 4 & \ 30 & 1.43 & \ 2.32 &
\ 20 & \hfil 0.4 & \hfil 0.6 & \hfil 1.15 \cr
\noalign{\smallskip}
\ 5 & \ 15 & 1.54 & \ 2.95 &
\ 5 & \hfil 0.4 & \hfil 0.6 & \hfil 1.07 \cr
\ 5 & \ 15 & 1.54 & \ 2.95 &
\ 10 & \hfil 0.0 & \hfil 0.5 & \hfil 0.99 \cr
\ 5 & \ 15 & 1.54 & \ 2.95 &
\ 10 & \hfil 0.4 & \hfil 0.6 & \hfil 1.08 \cr
\ 5 & \ 15 & 1.54 & \ 2.95 &
\ 10 & \hfil 0.2 & \hfil 0.8 & \hfil 1.03 \cr
%
%
\ 5 & \ 15 & 1.54 & \ 2.95 &
\ 20 & \hfil 0.4 & \hfil 0.6 & \hfil 1.07 \cr
\noalign{\smallskip}
\ 5 & \ 30 & 1.40 & \ 2.41 &
\ 5 & \hfil 0.4 & \hfil 0.6 & \hfil 1.20 \cr
\ 5 & \ 30 & 1.40 & \ 2.41 &
\ 10 & \hfil 0.0 & \hfil 0.5 & \hfil 1.01 \cr
\ 5 & \ 30 & 1.40 & \ 2.41 &
\ 10 & \hfil 0.4 & \hfil 0.6 & \hfil 1.14 \cr
\ 5 & \ 30 & 1.40 & \ 2.41 &
\ 10 & \hfil 0.2 & \hfil 0.8 & \hfil 1.07 \cr
%
%
\ 5 & \ 30 & 1.40 & \ 2.41 &
\ 20 & \hfil 0.4 & \hfil 0.6 & \hfil 1.11 \cr
\noalign{\smallskip}
\noalign{\hrule}
}}$$
\end{table}
%
\begin{table}
\caption{Same as Table 1 but for $E_{\rm ej} = 2 \times 10^{51}$ erg  
}
$$\vbox{\tabskip4pt \halign{#&#&#&#&#&#&#&# \cr
\noalign{\hrule}
\noalign{\smallskip}
\hfil $M_{\rm ej}$ & 
\, $n_{\rm H0}$ & \ $R_{\rm b}$ & \ $v_{\rm b}$
& \, $\alpha_{\rm clmp}$ & \ $q_0$ & \ $q_1$ & \ ${\cal A}$ \cr
(M$_\odot$) & (cm$^{-3}$)
& (pc) & $(10^3$ km/s) & & & \cr
\noalign{\smallskip}
\noalign{\hrule}
\noalign{\smallskip}
\noalign{\smallskip}
\ 2 & \  5 & 2.44 & \ 3.63 &
\ 5 & \hfil 0.4 & \hfil 0.6 & \hfil 1.19 \cr
\ 2 & \  5 & 2.44 & \ 3.63 &
\ 10 & \hfil 0.0 & \hfil 0.5 & \hfil 1.03 \cr
\ 2 & \  5 & 2.44 & \ 3.63 &
\ 10 & \hfil 0.4 & \hfil 0.6 & \hfil 1.33 \cr
\ 2 & \  5 & 2.44 & \ 3.63 &
\ 10 & \hfil 0.2 & \hfil 0.8 & \hfil 1.26 \cr
\ 2 & \  5 & 2.44 & \ 3.63 &
\ 20 & \hfil 0.4 & \hfil 0.6 & \hfil 1.51 \cr
\noalign{\smallskip}
\ 2 & \ 15 & 2.01 & \ 2.80 &
\ 5 & \hfil 0.4 & \hfil 0.6 & \hfil 1.53 \cr
\ 2 & \ 15 & 2.01 & \ 2.80 &
\ 10 & \hfil 0.0 & \hfil 0.5 & \hfil 1.14 \cr
\ 2 & \ 15 & 2.01 & \ 2.80 &
\ 10 & \hfil 0.4 & \hfil 0.6 & \hfil 2.07 \cr
\ 2 & \ 15 & 2.01 & \ 2.80 &
\ 10 & \hfil 0.2 & \hfil 0.8 & \hfil 1.53 \cr
\ 2 & \ 15 & 2.01 & \ 2.80 &
\ 20 & \hfil 0.4 & \hfil 0.6 & \hfil 2.03 \cr
\noalign{\smallskip}
\ 2 & \ 30 & 1.77 & \ 2.40 &
\ 5 & \hfil 0.4 & \hfil 0.6 & \hfil 2.12 \cr
\ 2 & \ 30 & 1.77 & \ 2.40 &
\ 10 & \hfil 0.0 & \hfil 0.5 & \hfil 1.36 \cr
\ 2 & \ 30 & 1.77 & \ 2.40 &
\ 10 & \hfil 0.4 & \hfil 0.6 & \hfil 2.67 \cr
\ 2 & \ 30 & 1.77 & \ 2.40 &
\ 10 & \hfil 0.2 & \hfil 0.8 & \hfil 1.76 \cr
\ 2 & \ 30 & 1.77 & \ 2.40 &
\ 20 & \hfil 0.4 & \hfil 0.6 & \hfil 2.11 \cr
\noalign{\smallskip}
\ 3 & \ 15 & 1.96 & \ 2.90 &
\ 5 & \hfil 0.4 & \hfil 0.6 & \hfil 1.47 \cr
\ 3 & \ 15 & 1.96 & \ 2.90 &
\ 10 & \hfil 0.0 & \hfil 0.5 & \hfil 1.09 \cr
\ 3 & \ 15 & 1.96 & \ 2.90 &
\ 10 & \hfil 0.4 & \hfil 0.6 & \hfil 1.74 \cr
\ 3 & \ 15 & 1.96 & \ 2.90 &
\ 10 & \hfil 0.2 & \hfil 0.8 & \hfil 1.33 \cr
%
%
\ 3 & \ 15 & 1.96 & \ 2.90 &
\ 20 & \hfil 0.4 & \hfil 0.6 & \hfil 1.54 \cr
\noalign{\smallskip}
\ 3 & \ 30 & 1.74 & \ 2.46 &
\ 5 & \hfil 0.4 & \hfil 0.6 & \hfil 2.04 \cr
\ 3 & \ 30 & 1.74 & \ 2.46 &
\ 10 & \hfil 0.0 & \hfil 0.5 & \hfil 1.30 \cr
\ 3 & \ 30 & 1.74 & \ 2.46 &
\ 10 & \hfil 0.4 & \hfil 0.6 & \hfil 2.13 \cr
\ 3 & \ 30 & 1.74 & \ 2.46 &
\ 10 & \hfil 0.2 & \hfil 0.8 & \hfil 1.53 \cr
%
%
\ 3 & \ 30 & 1.74 & \ 2.46 &
\ 20 & \hfil 0.4 & \hfil 0.6 & \hfil 1.57 \cr
\noalign{\smallskip}
\ 4 & \ 15 & 1.92 & \ 3.00 &
\ 5 & \hfil 0.4 & \hfil 0.6 & \hfil 1.36 \cr
\ 4 & \ 15 & 1.92 & \ 3.00 &
\ 10 & \hfil 0.0 & \hfil 0.5 & \hfil 1.04 \cr
\ 4 & \ 15 & 1.92 & \ 3.00 &
\ 10 & \hfil 0.4 & \hfil 0.6 & \hfil 1.48 \cr
\ 4 & \ 15 & 1.92 & \ 3.00 &
\ 10 & \hfil 0.2 & \hfil 0.8 & \hfil 1.20 \cr
%
%
\ 4 & \ 15 & 1.92 & \ 3.00 &
\ 20 & \hfil 0.4 & \hfil 0.6 & \hfil 1.31 \cr
\noalign{\smallskip}
\ 4 & \ 30 & 1.71 & \ 2.53 &
\ 5 & \hfil 0.4 & \hfil 0.6 & \hfil 1.82 \cr
\ 4 & \ 30 & 1.71 & \ 2.53 &
\ 10 & \hfil 0.0 & \hfil 0.5 & \hfil 1.20 \cr
\ 4 & \ 30 & 1.71 & \ 2.53 &
\ 10 & \hfil 0.4 & \hfil 0.6 & \hfil 1.74 \cr
\ 4 & \ 30 & 1.71 & \ 2.53 &
\ 10 & \hfil 0.2 & \hfil 0.8 & \hfil 1.36 \cr
%
%
\ 4 & \ 30 & 1.71 & \ 2.53 &
\ 20 & \hfil 0.4 & \hfil 0.6 & \hfil 1.35 \cr
\noalign{\smallskip}
\ 5 & \ 15 & 1.87 & \ 3.10 &
\ 5 & \hfil 0.4 & \hfil 0.6 & \hfil 1.25 \cr
\ 5 & \ 15 & 1.87 & \ 3.10 &
\ 10 & \hfil 0.0 & \hfil 0.5 & \hfil 1.01 \cr
\ 5 & \ 15 & 1.87 & \ 3.10 &
\ 10 & \hfil 0.4 & \hfil 0.6 & \hfil 1.31 \cr
\ 5 & \ 15 & 1.87 & \ 3.10 &
\ 10 & \hfil 0.2 & \hfil 0.8 & \hfil 1.12 \cr
%
%
\ 5 & \ 15 & 1.87 & \ 3.10 &
\ 20 & \hfil 0.4 & \hfil 0.6 & \hfil 1.20 \cr
\noalign{\smallskip}
\ 5 & \ 30 & 1.68 & \ 2.59 &
\ 5 & \hfil 0.4 & \hfil 0.6 & \hfil 1.61 \cr
\ 5 & \ 30 & 1.68 & \ 2.59 &
\ 10 & \hfil 0.0 & \hfil 0.5 & \hfil 1.11 \cr
\ 5 & \ 30 & 1.68 & \ 2.59 &
\ 10 & \hfil 0.4 & \hfil 0.6 & \hfil 1.49 \cr
\ 5 & \ 30 & 1.68 & \ 2.59 &
\ 10 & \hfil 0.2 & \hfil 0.8 & \hfil 1.23 \cr
%
%
\ 5 & \ 30 & 1.68 & \ 2.59 &
\ 20 & \hfil 0.4 & \hfil 0.6 & \hfil 1.24 \cr
\noalign{\smallskip}
\noalign{\hrule}
}}$$
\end{table}
%
\begin{table}
\caption{Same as Table 1 but for $E_{\rm ej} = 3 \times 10^{51}$ erg  
}
$$\vbox{\tabskip4pt \halign{#&#&#&#&#&#&#&# \cr
\noalign{\hrule}
\noalign{\smallskip}
\hfil $M_{\rm ej}$ & 
\, $n_{\rm H0}$ & \ $R_{\rm b}$ & \ $v_{\rm b}$
& \, $\alpha_{\rm clmp}$ & \ $q_0$ & \ $q_1$ & \ ${\cal A}$ \cr
(M$_\odot$) & (cm$^{-3}$)
& (pc) & $(10^3$ km/s) & & & \cr
\noalign{\smallskip}
\noalign{\hrule}
\noalign{\smallskip}
\noalign{\smallskip}
\ 2 & \  5 & 2.68 & \ 3.84 &
\ 5 & \hfil 0.4 & \hfil 0.6 & \hfil 1.22 \cr
\ 2 & \  5 & 2.68 & \ 3.84 &
\ 10 & \hfil 0.0 & \hfil 0.5 & \hfil 1.04 \cr
\ 2 & \  5 & 2.68 & \ 3.84 &
\ 10 & \hfil 0.4 & \hfil 0.6 & \hfil 1.36 \cr
\ 2 & \  5 & 2.68 & \ 3.84 &
\ 10 & \hfil 0.2 & \hfil 0.8 & \hfil 1.36 \cr
\ 2 & \  5 & 2.68 & \ 3.84 &
\ 20 & \hfil 0.4 & \hfil 0.6 & \hfil 1.70 \cr
\noalign{\smallskip}
\ 2 & \ 15 & 2.20 & \ 3.00 &
\ 5 & \hfil 0.4 & \hfil 0.6 & \hfil 1.44 \cr
\ 2 & \ 15 & 2.20 & \ 3.00 &
\ 10 & \hfil 0.0 & \hfil 0.5 & \hfil 1.09 \cr
\ 2 & \ 15 & 2.20 & \ 3.00 &
\ 10 & \hfil 0.4 & \hfil 0.6 & \hfil 2.13 \cr
\ 2 & \ 15 & 2.20 & \ 3.00 &
\ 10 & \hfil 0.2 & \hfil 0.8 & \hfil 1.66 \cr
\ 2 & \ 15 & 2.20 & \ 3.00 &
\ 20 & \hfil 0.4 & \hfil 0.6 & \hfil 2.57 \cr
\noalign{\smallskip}
\ 2 & \ 30 & 1.93 & \ 2.57 &
\ 5 & \hfil 0.4 & \hfil 0.6 & \hfil 1.87 \cr
\ 2 & \ 30 & 1.93 & \ 2.57 &
\ 10 & \hfil 0.0 & \hfil 0.5 & \hfil 1.22 \cr
\ 2 & \ 30 & 1.93 & \ 2.57 &
\ 10 & \hfil 0.4 & \hfil 0.6 & \hfil 2.85 \cr
\ 2 & \ 30 & 1.93 & \ 2.57 &
\ 10 & \hfil 0.2 & \hfil 0.8 & \hfil 1.87 \cr
\ 2 & \ 30 & 1.93 & \ 2.57 &
\ 20 & \hfil 0.4 & \hfil 0.6 & \hfil 2.77 \cr
\noalign{\smallskip}
\ 3 & \ 15 & 2.16 & \ 3.08 &
\ 5 & \hfil 0.4 & \hfil 0.6 & \hfil 1.47 \cr
\ 3 & \ 15 & 2.16 & \ 3.08 &
\ 10 & \hfil 0.0 & \hfil 0.5 & \hfil 1.09 \cr
\ 3 & \ 15 & 2.16 & \ 3.08 &
\ 10 & \hfil 0.4 & \hfil 0.6 & \hfil 2.01 \cr
\ 3 & \ 15 & 2.16 & \ 3.08 &
\ 10 & \hfil 0.2 & \hfil 0.8 & \hfil 1.49 \cr
%
%
\ 3 & \ 15 & 2.16 & \ 3.08 &
\ 20 & \hfil 0.4 & \hfil 0.6 & \hfil 2.01 \cr
\noalign{\smallskip}
\ 3 & \ 30 & 1.90 & \ 2.63 &
\ 5 & \hfil 0.4 & \hfil 0.6 & \hfil 2.07 \cr
\ 3 & \ 30 & 1.90 & \ 2.63 &
\ 10 & \hfil 0.0 & \hfil 0.5 & \hfil 1.30 \cr
\ 3 & \ 30 & 1.90 & \ 2.63 &
\ 10 & \hfil 0.4 & \hfil 0.6 & \hfil 2.63 \cr
\ 3 & \ 30 & 1.90 & \ 2.63 &
\ 10 & \hfil 0.2 & \hfil 0.8 & \hfil 1.72 \cr
%
%
\ 3 & \ 30 & 1.90 & \ 2.63 &
\ 20 & \hfil 0.4 & \hfil 0.6 & \hfil 2.12 \cr
\noalign{\smallskip}
\ 4 & \ 15 & 2.12 & \ 3.16 &
\ 5 & \hfil 0.4 & \hfil 0.6 & \hfil 1.42 \cr
\ 4 & \ 15 & 2.12 & \ 3.16 &
\ 10 & \hfil 0.0 & \hfil 0.5 & \hfil 1.06 \cr
\ 4 & \ 15 & 2.12 & \ 3.16 &
\ 10 & \hfil 0.4 & \hfil 0.6 & \hfil 1.78 \cr
\ 4 & \ 15 & 2.12 & \ 3.16 &
\ 10 & \hfil 0.2 & \hfil 0.8 & \hfil 1.34 \cr
%
%
\ 4 & \ 15 & 2.12 & \ 3.16 &
\ 20 & \hfil 0.4 & \hfil 0.6 & \hfil 1.64 \cr
\noalign{\smallskip}
\ 4 & \ 30 & 1.88 & \ 2.68 &
\ 5 & \hfil 0.4 & \hfil 0.6 & \hfil 2.02 \cr
\ 4 & \ 30 & 1.88 & \ 2.68 &
\ 10 & \hfil 0.0 & \hfil 0.5 & \hfil 1.26 \cr
\ 4 & \ 30 & 1.88 & \ 2.68 &
\ 10 & \hfil 0.4 & \hfil 0.6 & \hfil 2.26 \cr
\ 4 & \ 30 & 1.88 & \ 2.68 &
\ 10 & \hfil 0.2 & \hfil 0.8 & \hfil 1.55 \cr
%
%
\ 4 & \ 30 & 1.88 & \ 2.68 &
\ 20 & \hfil 0.4 & \hfil 0.6 & \hfil 1.71 \cr
\noalign{\smallskip}
\ 5 & \ 15 & 2.08 & \ 3.24 &
\ 5 & \hfil 0.4 & \hfil 0.6 & \hfil 1.35 \cr
\ 5 & \ 15 & 2.08 & \ 3.24 &
\ 10 & \hfil 0.0 & \hfil 0.5 & \hfil 1.02 \cr
\ 5 & \ 15 & 2.08 & \ 3.24 &
\ 10 & \hfil 0.4 & \hfil 0.6 & \hfil 1.57 \cr
\ 5 & \ 15 & 2.08 & \ 3.24 &
\ 10 & \hfil 0.2 & \hfil 0.8 & \hfil 1.24 \cr
%
%
\ 5 & \ 15 & 2.08 & \ 3.24 &
\ 20 & \hfil 0.4 & \hfil 0.6 & \hfil 1.42 \cr
\noalign{\smallskip}
\ 5 & \ 30 & 1.85 & \ 2.74 &
\ 5 & \hfil 0.4 & \hfil 0.6 & \hfil 1.88 \cr
\ 5 & \ 30 & 1.85 & \ 2.74 &
\ 10 & \hfil 0.0 & \hfil 0.5 & \hfil 1.19 \cr
\ 5 & \ 30 & 1.85 & \ 2.74 &
\ 10 & \hfil 0.4 & \hfil 0.6 & \hfil 1.94 \cr
\ 5 & \ 30 & 1.85 & \ 2.74 &
\ 10 & \hfil 0.2 & \hfil 0.8 & \hfil 1.41 \cr
%
%
\ 5 & \ 30 & 1.85 & \ 2.74 &
\ 20 & \hfil 0.4 & \hfil 0.6 & \hfil 1.47 \cr
\noalign{\smallskip}
\noalign{\hrule}
}}$$
\end{table}

For certain combinations of the characterizing 
parameters of the 
supernova remnant model, $E_{\rm ej}$, $M_{\rm ej}$
and $n_{\rm H0}$, the delayed decay of ionized \Ti\ 
can have a sizable effect on the $\gamma$-ray activity of Cas~A which 
is observable at the present age of about 320 years. In Tables~1--3
the ratio ${\cal A}$ 
denotes the currently observable activity according
to our model, normalized to the expected activity if ionization 
effects are not taken into account. The results depend on the assumed
density enhancement $\alpha_{\rm clmp}$ in the clumps relative to the 
density of the homogeneous ejecta, and on the location of the clumps 
within a shell in the expanding ejecta bounded by the lower and upper 
mass coordinates $q_0$ and $q_1$, respectively. 
In addition, in Tables~1--3
numbers are given for the blast-wave radius $R_{\rm b}$
and the blast-wave
velocity $v_{\rm b}$ as predicted by the McKee \& Truelove (1995) model at the 
current age of the Cas~A supernova remnant. 

The ratio
${\cal A}$ exhibits the following tendencies. For fixed explosion energy 
and ejecta mass, ${\cal A}$ 
increases with the density of the circumstellar
gas of the supernova remnant because the expansion and dilution of the
ejecta are slowed down for higher $n_{\rm H0}$ (i.e., $v_{\rm b}$ and
$R_{\rm b}$ are smaller at the same time). An increase of the ratio
${\cal A}$ can also be seen when the explosion energy becomes larger
but all other parameters are kept constant. The opposite trend is visible
if the explosion energy and ambient density are fixed, in which case
the ratio ${\cal A}$ decreases with larger ejecta mass.
Although the blast-wave velocity at the present time is higher,
it was lower early after the explosion and therefore the blast-wave
radius is smaller for the larger ejecta mass. 

It is not easy to give simple explanations for these trends.
The ratio ${\cal A}$, which scales with the current abundance of
\Ti\  and its effective decay rate, reflects the whole time
evolution of the remnant and contains contributions from all parts of the
shocked ejecta with their different ionization histories. The nonlinear 
dependence of ${\cal A}$
on the remnant parameters is a result of the interplay between a number of
effects. 
For example, the post-shock density and temperature in the ejecta,
in particular in the clumps, are important for the degree of ionization.
The time when the reverse shock hits the clumps determines the 
time left for the equilibration between electrons and ions and for
the duration of the delay of the \Ti\ decay. 
This delay lasts from the moment of nearly complete ionization
until recombination takes place again.

In order to obtain a large effect on the \Ti\  radioactivity that is 
measured in Cas~A presently, a high post-shock
density and temperature are favorable for efficient ionization 
of \Ti. On the other hand, the present-day $\gamma$-ray emission of  
Cas~A means that the temperature must not have decreased
 too slowly and the 
density not too rapidly. If that happened, \Ti\ would be hindered  
from recombination and the observed $\gamma$-ray emission could not be 
explained. The maximum effect from the delayed decay of ionized 
\Ti\ is found if most of the \Ti-carrying clumps have
been mixed  roughly half-way into the ejecta,
 i.e., if they are assumed to 
be located in the $q$-interval between 0.4 and 0.6 (compare
Fig.~2), and if the density enhancement in the clumps
is around 10 for the higher explosion energies of 
$E_{\rm ej} = 2\times 10^{51}\,{\rm erg}$ and
 $3\times 10^{51}\,{\rm erg}$
(Tables~2 and 3, respectively). In case of the lowest considered 
value of
the explosion energy, $E_{\rm ej} = 1\times 10^{51}\,{\rm erg}$ 
(Table~1),
a clump over-density  of about 5 seems
 preferable because of the higher
density of the more slowly expanding ejecta. 
Higher than optimum $\alpha_{\rm clmp}$-values do not ensure
rapid and strong ionization, whereas smaller than optimum
$\alpha_{\rm clmp}$-values do not allow for quick recombination
so that the decay rate stays low.
If \Ti\  is located very
far out in the remnant (large $q$), the temperature
 behind the reverse shock 
is too low for efficient ionization 
(see Fig.~1). If \Ti\      
sits deep inside the ejecta (small $q$), the clumps 
have been reached by the
reverse shock not sufficiently long ago so as to show a 
significant effect from \Ti\ 
ionization.

We found maximum values of ${\cal A}$ between roughly 
1.5 and 2.5 for
a large number of different combinations of remnant 
parameters and assumed
clump locations and density enhancement factors. This means that the 
present-day \Ti\  radioactivity of Cas~A could
 be 50\% up to more than a 
factor of 2 higher for  a certain amount of \Ti,
if the decay half-life of \Ti\  was stretched by 
ionization effects. Or, reversely, the current $\gamma$-ray emission of
Cas~A due to \Ti\ decay might be explained with a significantly lower
production of this nucleus during the supernova explosion. Figure~4 shows
that the relative effect from the retardation of the \Ti\ decay
will increase with
the age of the remnant.

For the different considered combinations of parameter values,
Cas~A has reached
slightly different stages of the remnant evolution around the
transition time from the
ejecta-dominated to the Sedov-Taylor phase. The tabulated 
blast-wave radius
and velocity can be compared with observations which give values for
the circumstellar density of about 20 hydrogen atoms per cm$^3$,
an estimated 
ejecta mass of about $(2 \sim 5)\,$M$_{\odot}$
 and a
blast-wave radius of 2--3~pc at a distance of 3.4~kpc 
(see references given in Sect.~2).

\section{Summary}

We examined the effects of the reverse shock 
on the \Ti\ $\beta$-decay rate in young supernova remnants.
For this purpose, we employed the  analytic remnant
 model of McKee \& Truelove
(1995), which was used to describe the hydrodynamic evolution.
We assumed that \Ti\ is carried by $^{56}$Fe-dominated, dense clumps
which are mixed into the otherwise homogeneous supernova ejecta
and account for a minor fraction of the ejecta mass.
Strong ionization of \Ti\ due to the heating by the reverse shock
and a sizable influence on the \Ti\ decay is obtained if the clumps
are assumed to be located at intermediate values of the radial
mass coordinate $q$.
The observationally important change of the \Ti\ decay activity,
however, depends 
also on the values of the remnant parameters,
namely  
the explosion energy,
the ejecta mass, the density of the ambient medium and the assumed 
overdensity factor of the clumps relative to the embedding
homogeneous ejecta.
 
%
\begin{figure}
\epsfxsize=0.99\hsize
\epsffile{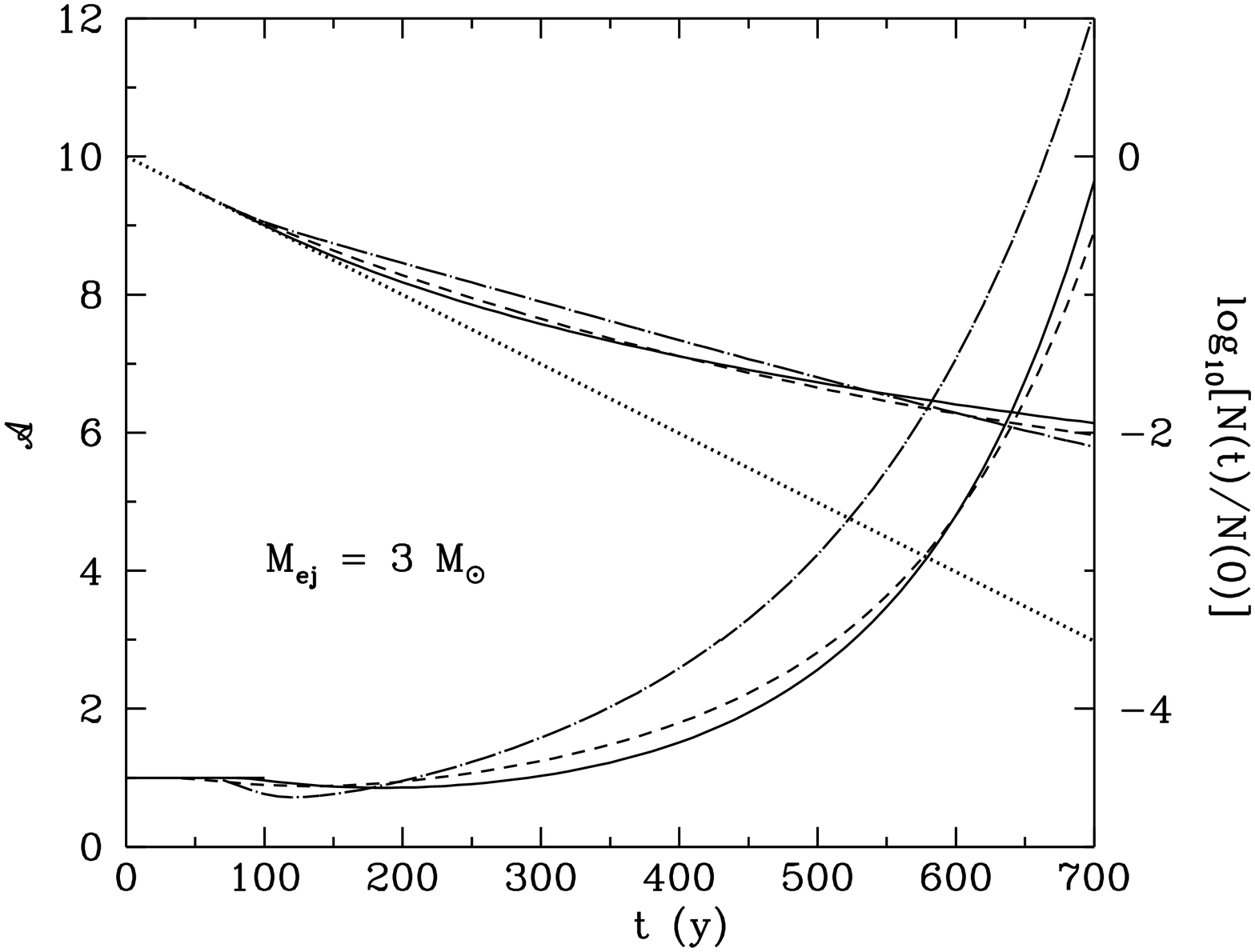}
\epsfxsize=0.99\hsize
\epsffile{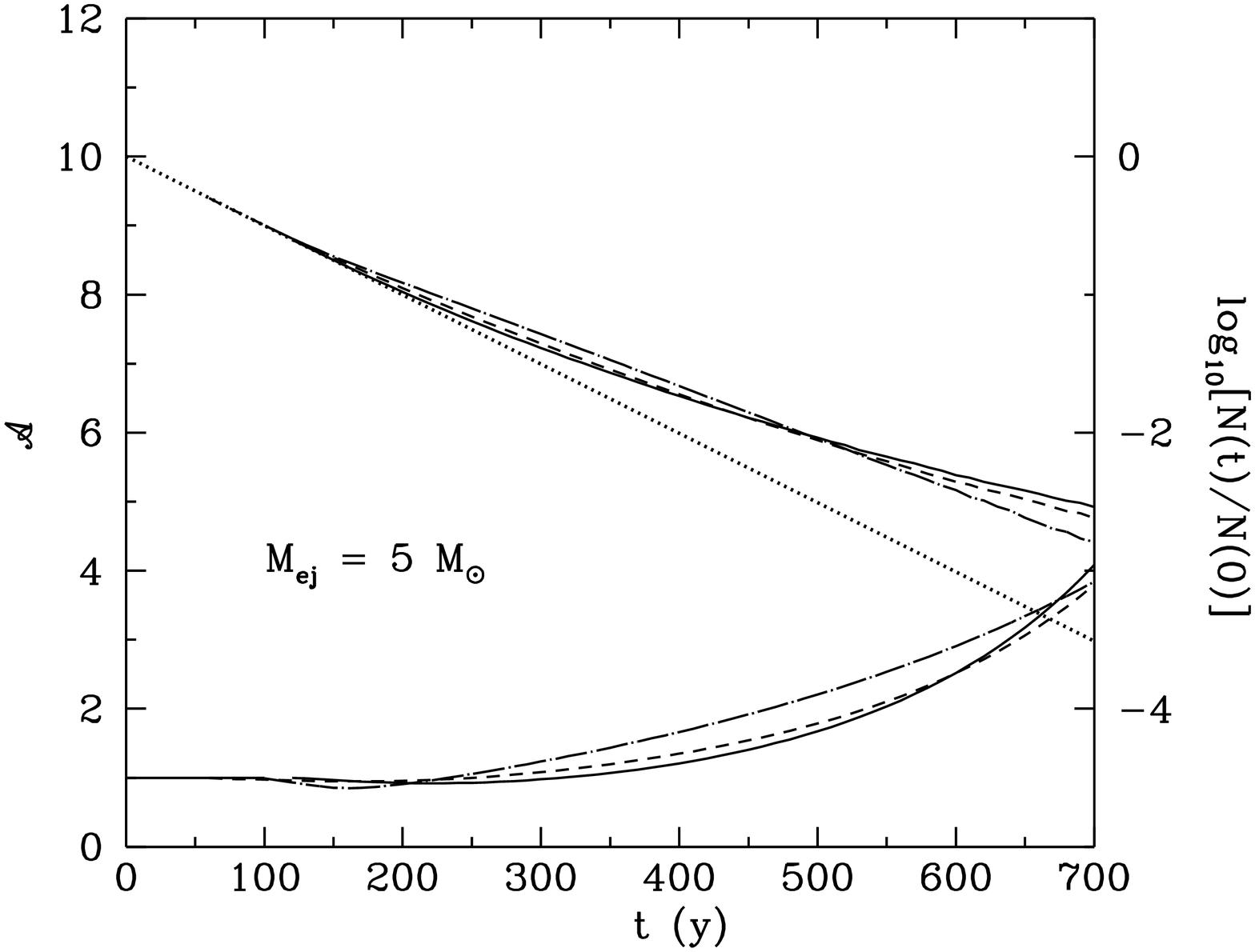}
\caption{The decrease with time 
of the amount of  \Ti\ in the supernova remnant, $N(t)$,
by $\beta$-decays, starting from the initial amount  $N(0)$
({\it right scale} in logarithm), and the corresponding,
mostly increasing activity ratio ${\cal A}$ ({\it left scale}).
Values of 3 M$_\odot$ ({\it upper panel}) and
5 M$_\odot$ ({\it lower}) are used for $M_{\rm ej}$, while
$E_{\rm ej} = 2 \times 10^{51}$ erg, $n_{\rm H0} = 15$ cm$^{-3}$
and $\alpha_{\rm clmp} = 10$ are adopted in both cases.
The different lines show results obtained for different
assumptions of the radial locations of 
the clumps:
$q = [q_0, q_1]$ = [0.0, 0.5] ({\it solid line}), 
[0.4, 0.6] ({\it dash-dotted line}), and [0.2, 0.8] ({\it dashed line}).
The exponential decrease of  the \Ti\ abundance with the laboratory
rate is plotted by the {\it dotted line}
}
\end{figure}
We applied our model to the case of Cas~A,
a young supernova remnant with an age of about 320 years.
Observational data are considered to constrain the parameter
space which was explored.
We found that under certain conditions the ionization of \Ti\ 
and the corresponding delay of its decay can yield an up to three
times higher \Ti\ activity at the present time than predicted
on the grounds of the laboratory decay rate.
This effect 
is large enough to
reduce the apparent discrepancies between the \Ti\ production
in the explosion as
inferred from the COMPTEL $\gamma$-line measurements
and the theoretical expectations 
from the current supernova nucleosynthesis models.
 
We emphasize again that  to get an enhanced \Ti\ decay
activity we had to assume \Ti\ to be associated with  
inhomogeneities in the ejecta which contain the Fe-group
elements. Hydrodynamical models have so far not been able to
predict the fraction of Fe and \Ti\ in such clumpy structures,
the overdensity of the clumps relative to the surrounding
material at the considered age of  a supernova remnant,
the dynamics of the clumps, and their size and distribution.
Multi-dimensional supernova models are called for, which connect the  
very early phase of the explosion with the remnant evolution
a few hundred years later.
Self-consistent multi-dimensional simulations
have yet to
be carried out for an improved theoretical picture
of the explosive \Ti\ production in supernovae. 
All published theoretical yields have so far only been obtained
with ad-hoc assumptions, most often 
by spherically symmetric models and/or simulations of the
supernova explosions which were started from artificial
initial conditions (``piston'' models)
rather than self-consistent situations.
We add that such calculations also
suffer from uncertainties concerning the
nuclear reaction rates which  are important to determine the 
\Ti\ yield (The et al. 1998).
 
On the other hand, more detailed X-ray and $\gamma$-ray
observations with a better spatial (and energy) resolution
are desirable to confirm or reject the implications of the 
described model, e.g., the existence of  Fe-containing knots
with temperatures in the approximate range of $(1 \sim 5)$ keV 
(cf. Tsunemi 1997). 
The current angular resolution of the ASCA/SIS observations, about
one arcmin, is insufficient to measure the 
thermodynamic properties of individual clumps in the ejecta of Cas~A.
Only information is available for the average electron temperature in
the matter that is heated by the forward and reverse shocks. Comparing
the observations with 
predictions from the remnant model used in this paper
would therefore require monitoring of the thermodynamic history of
the homogeneous ejecta.
For this, one would need to compute 
the thermal state of the electrons, and would have to
specify 
the unknown composition
of the ejecta,
which is very uncertain because it depends on the 
type of progenitor and explosion.
The described investigations of 
$^{44}$Ti decay in supernova remnants could be done without such
additional assumptions, but the models do not provide
a description of the thermodynamic state of the remnant, which 
allows for a direct comparison with current observational data.
 
Finally, we mention that we applied our analysis 
also to the new supernova remnant, RX~J0852.0-4622,
discovered 
in the direction of the Vela remnant by Aschenbach (1998),
 in which
a source of \Ti\ line emission has been detected
(Iyudin et al. 1998).
However, for the estimated low density of the ambient medium
of less than 
$0.04 (d / 500 {\rm pc})^{-0.5}$ cm$^{-3}$ with a distance $d$
to the remnant (Aschenbach 1998),
we found that the reverse shock does not heat the ejecta
to sufficiently high temperatures to ionize \Ti.

{\acknowledgements{
We thank H.~Tsunemi, J.~Vink, E.~Miyata and K.~Nomoto for 
helpful information and discussions.
This work was supported by the SFB-375 ``Astroparticle
Physics'' of the Deutsche Forschungsgemeinschaft.
}}
 
\vskip12pt
\noindent
{\bf Appendix A: reaction rates and related uncertainties}
%
\vskip8pt
\noindent
We estimate here the $\beta$-decay, ionization and recombination rates,
which appear in Eq.~(7).
We also evaluate 
their uncertainties in relation to the results presented in Sect.~4.
 
\vskip10pt
{\it The orbital-electron capture rate of $^{44}${\rm Ti}}:
We first formulate the $\beta$-decay rate of (near) neutral \Ti. 
The corresponding half-life $t_{1/2}$
has been well determined to be
60 y (Ahmad et al. 1998; G\"orres et al. 1998;  Norman et al. 1998;
Wietfeldt et al. 1999).
The \Ti\ $\beta$-decay is by orbital-electron capture
almost exclusively
to the 146 keV excited ($0^-$) state of $^{44}$Sc (Endt 1998), 
the corresponding $Q$-value being 266 keV.
We thus write the laboratory rate, $\lambda_{\rm lab} = 
{\rm ln}~2/t_{1/2}$, as
%
\begin{equation}
   \lambda_{\rm lab}  =  \lambda_{\rm K} +
    \lambda_{\rm LI} + ....
\end{equation}
\noindent
where $\lambda_{\rm K}, \lambda_{\rm LI}$ etc. refer to the
rates for partial 
captures of K (1s$_\frac{1}{2}$), LI (2s$_\frac{1}{2}$)
electrons etc.
In the ``normal'' approximation (e.g. Konopinski \& Rose 1965), 
$\lambda_{\rm K}$ for example reads
%
\begin{equation}
   \lambda_{\rm K} =  \frac{G_{\rm A}^2}{4\pi^2}
   | M_0 |^2  (Q + B_{\rm K} )^2  g_{\rm K}^2 \, , 
\end{equation}
\noindent
for the transition of our current interest,
where $G_{\rm A}$ is the axial-vector coupling constant of
weak interaction,
$M_0$ is a linear combination of two major nuclear
matrix elements in the rank-0 first-forbidden ($0^+ \rightarrow 0^-$)
transition,
$Q$ is the $Q$-value, and
$B_{\rm K}$ is the binding energy of the K-shell.
Thus, $q_\nu \equiv Q + B_{\rm K}$ is the energy
available to the emitted neutrino.
(The nuclear recoil energy is very small and is ignored here.)
Finally, $g_{\rm K}$ is the larger component of the radial
wave-functions of the K-electron, which is evaluated 
at the nuclear radius.
 
Since $Q$ is much larger than the binding energies (about 5 keV
for K-electrons),
the relative ratios of the partial decay rates
become nearly equal to those of the
respective wave-functions squared.
Considering in addition that $\alpha Z $ is  not very large,
we have 
%
\begin{equation}
  \lambda_{\rm LI} \approx \frac{1}{8} \lambda_{\rm K}\, ,
\end{equation}
since $g_{\rm LI}^2
\approx g_{\rm K}^2/8$ in 
the low-$\alpha Z$ approximation.
With the use of the same approximation, one finds that
the capture rates of the higher-shell electrons are small.
 
Correspondingly, 
the orbital-electron capture rates of \Ti\ with $i$ orbital
electrons [i.e., (22-$i$)-times ionized]
are approximately given by
%
\begin{equation}
  \lambda_{\beta,i} \approx  p_{i,{\rm K}} \lambda_{i,{\rm K}}
  + p_{i,{\rm LI}} \lambda_{i,{\rm LI}}\, ,
\end{equation}
\noindent
where $p_{i,{\rm K}}$ and $p_{i,{\rm LI}}$ are the 
occupancies ([0,1]) of the K  and L$_{\rm I}$
orbits, respectively, and $\lambda_{i,{\rm K}}$ and 
$\lambda_{i,{\rm LI}}$ are the
 corresponding capture rates for the {\it filled} shells. 
We approximate the partial capture rates by
%
\begin{equation}
\lambda_{i,{\rm K}} \approx \frac{8}{9} \lambda_{i,{\rm K+LI}} \approx
 \frac{8}{9} \lambda_{\rm lab},\  
\lambda_{i,{\rm LI}} \approx \frac{1}{9} \lambda_{i,{\rm K+LI}} \approx
 \frac{1}{9} \lambda_{\rm lab}\, ,
\end{equation} 
\noindent 
so that the total (effective) decay rate reads
%
\begin{equation}
 \lambda_{\rm eff} \approx \Bigl( \frac{4}{9} n_1 + \frac{8}{9} n_2 
 + \frac{17}{18} n_3 + \sum_{i \geq 4} n_i \Bigr) \lambda_{\rm lab}
\, {\bigg /}\,
 \sum_{i \geq 0} n_i \, ,
\end{equation}
\noindent
where $n_i$ is the number abundance of \Ti$^{+22-i}$.
 
The above procedure is quite well suited for dealing with  the decays of 
highly-ionized \Ti.
In fact, our calculations based on a relativistic mean field method 
(Libermann et al. 1971) show that
the overall errors associated with those approximations, including
the neglect of the decrease of screening and
the slightly altered energetics along with ionization,
are small enough (less than 10 \%)
to be ignored in the present work.
 
\vskip10pt
{\it The ionization rates of 
$^{44}${\rm Ti} and $^{56}${\rm Fe}}:
The rates of electron-induced ionization are given by
%
\begin{equation}
 \lambda_{{\rm ion},i} = n_{\rm e} \sum_{i \geq 1} 
\Bigl( p_{i,{\rm K}} \left\langle \sigma_{{\rm ion},i}^{\rm (K)} v
\right\rangle +
     p_{i,{\rm LI}} \left\langle \sigma_{{\rm ion},i}^{\rm (LI)} v
\right\rangle +...
 ... \Bigr)\, ,
\end{equation}
\noindent
where $n_{\rm e}$ is the electron number density and
  $\left\langle \sigma_{{\rm ion},i} v \right\rangle $ is
the Maxwellian-average of the cross section
(times the electron velocity $v$)
for ionization of filled K-, LI-...
shell electrons.
As in Eq.~(16), $p_{i,{\rm K}}$ and $p_{i,{\rm LI}}$ stand for
occupancies of the K- and L$_{\rm I}$ shells in the $i$-th ion. 
Our estimates of $\sigma_{{\rm ion},i}$ rely on the experimental data
on K-shell ionization by electrons in the relevant energy range of
$6 - 50$ keV (Long et al. 1990).
 
We start with the classical Bethe-Mott-Massey formula
(e.g., Powell 1976), which
for the K-shell ionization from an ion with $i$ bound electrons reads
%
\begin{equation}
    \sigma_{{\rm ion},i}^{\rm (K)} = 
[ \pi {\rm e}^4 / (E B_{i,\rm K}) ] 
  Z_{\rm K} b_{i,{\rm K}} {\rm ln}[4 c_{i,{\rm K}} E /
  B_{i,{\rm K}}]\, ,
\end{equation}
\noindent
where $B_{i,{\rm K}}$  is the K-binding energy (taken positive),
$E  (\geq B_{i,{\rm K}})$  is the incident electron energy, 
and $Z_{\rm K} = 2$ is the number of K-electrons, and
we  treat $b_{i,{\rm K}}$ and $c_{i,{\rm K}}$ as adjustable parameters.  
It turns out that the existing experimental 
data for  K-ionization of Ti by electrons
in the relevant $E$ range of 6 - 50 keV (Long et al. 1990)
can be well reproduced by the choice of
%
\begin{equation}
 1 / c_{22,{\rm K}} \approx 
a + ( 4 - a ) {\rm exp}[ d ( E / B_{22,{\rm K}} - 1 )]\, ,
\end{equation}
with $ a = 0.55$, $d = 0.45$, and $b_{22,{\rm K}} = 0.445.$
For simplicity, we use the same parameter values for the higher
shells as well as for highly-ionized cases.
Aside from the trivial replacements of $Z_{\rm K}$ by
$Z_{\rm LI}$ etc., we use the binding energies computed for each
ion by the relativistic formalism, because the rates are sensitive to
those thresholds. For \Ti, their values in keV are, e.g.,
$B_{1,{\rm K}} = 6.6, B_{2,{\rm K}} = 6.2, B_{3,{\rm K}} = 6.1,
B_{3,{\rm LI}} = 1.4, B_{4,{\rm K}} = 5.9, B_{4,{\rm LI}} = 1.3,$
in contrast to 
$B_{22,{\rm K}} = 5.0, B_{22,{\rm LI}} = 0.56.$
 
The same parameter values also reproduce the 
K-shell ionization cross-section data for Ni (Long et al. 1990)
 within the 
experimental uncertainties, while slightly (up
 to 20 \%) over-estimating the 
limited data points for Mn.
When considered for Fe,  this order of the uncertainty of
the ionization rates influences the radioactivity under
consideration in this work only at a negligible level.
 
\vskip10pt
{\it The recombination rates of 
$^{44}${\rm Ti} and $^{56}${\rm Fe}}:
The recombination rates are given by
%
\begin{equation}
 \lambda_{{\rm rec},i} = n_{\rm e} \sum_{i \geq 0} 
\Bigl( q_{i,{\rm K}} \left\langle \sigma_{{\rm rec},i}^{\rm (K)} v
\right\rangle  + 
     q_{i,{\rm LI}} \left\langle \sigma_{{\rm rec},i}^{\rm (LI)} v
\right\rangle  + 
 ... \Bigr)\, ,
\end{equation}
\noindent
where $q_{i} \equiv 1 - p_{i}$ are the vacancies of the
respective shell prior to the recombination.
We adopt the well-known formulae (Stobbe 1930) for
the recombination cross-sections in the non-relativistic approximation 
for a pure Coulomb field.
They are expressed in analytic forms in terms of the ratios
of the incident electron energies ($E \geq 0$)
 to the binding energies of 
respective shells.
The relativistic and screening effects can roughly be estimated with the
use of effective charges, which turn out to be largely insignificant
for the purpose of the present work.


%
\end{document}